\documentclass[11pt,3p,review,  authoryear,reqno]{elsarticle}
\journal{Energy Conversion and Management}
\makeatletter
\def\ps@pprintTitle{%
 \let\@oddhead\@empty
 \let\@evenhead\@empty
 \def\@oddfoot{}%
 \let\@evenfoot\@oddfoot}
\makeatother

\usepackage{setspace}
\usepackage{graphicx}
\usepackage{comment}
\usepackage{natbib}
\setcitestyle{square,numbers}
\usepackage[bottom]{footmisc}
\usepackage[]{algorithm2e}

\biboptions{longnamesfirst}
\usepackage{float, subcaption}
\usepackage{amssymb,amsmath,multicol, multirow}
\usepackage[english]{babel} 
\usepackage{siunitx}  

\usepackage[font=normalsize]{caption}
\usepackage{url}
\usepackage{textcomp}

\usepackage{algpseudocode}
\usepackage{adjustbox}
\usepackage[table,xcdraw]{xcolor}
\usepackage[toc,page]{appendix}
\usepackage{todonotes}

\usepackage{color}

\makeatletter
\newcommand{\thickhline}{%
    \noalign {\ifnum 0=`}\fi \hrule height 2pt
    \futurelet \reserved@a \@xhline
}
\newcolumntype{"}{@{\hskip\tabcolsep\vrule width 2pt\hskip\tabcolsep}}
\makeatother

\usepackage{soul}

\begin{document}


\begin{frontmatter}
\title{Transfer Learning for Electricity Price Forecasting}
\author[ITU]{Salih Gunduz\corref{cor1}}
\ead{gunduzs20@itu.edu.tr}
\author[BAU]{Umut Ugurlu }
\author[ITU,KCL]{Ilkay Oksuz}

\cortext[cor1]{ Computer Engineering Department, Istanbul Technical University, Istanbul, Turkey}
\address[ITU]{Computer Engineering Department, Istanbul Technical University, Istanbul, Turkey}
\address[BAU]{Management Department, Bahcesehir University, Istanbul, Turkey}
\address[KCL]{School of Biomedical Engineering \& Imaging Sciences, King\textquotesingle s College London, U.K}

\begin{abstract}

Electricity price forecasting is an essential task in all the deregulated markets of the world. The accurate prediction of the day-ahead electricity prices is an active research field and available data from various markets can be used as an input for forecasting. A collection of models have been proposed for this task, but the fundamental question on how to use the available big data is often neglected. In this paper, we propose to use transfer learning as a tool for utilizing information from other electricity price markets for forecasting. We pre-train a neural network model on source markets and finally do a fine-tuning for the target market. Moreover, we test different ways to use the rich input data from various electricity price markets. Our experiments on four different day-ahead markets indicate that transfer learning improves the electricity price forecasting performance in a statistically significant manner. Furthermore, we compare our results with state-of-the-art methods in a rolling window scheme to demonstrate the performance of the transfer learning approach.

\end{abstract}
\end{frontmatter}


\section{Introduction}
\label{sec:introduction}

Forecasting electricity prices accurately has been a major task since the establishment of the liberalized electricity markets. The players on both sides of the market aim to forecast the prices accurately for generation and profit optimization. The task has been studied individually in different markets. However, learning inter-dependent information in between different markets is an under-studied field.

Recently, deep learning methods have showcased superior performance in predicting electricity prices \cite{lago2018forecasting, ugurlu2018}. Most of the literature on the application of neural networks for electricity price forecasting has relied on single market data and available large amounts of data from different markets have not been utilized.

Transfer Learning is a major tool to improve performance on many tasks such as image classification, machine translation, and speech recognition. It is a machine learning method, where a model trained for a specific task is the initialization point for a model training on a second task. In this paper, we utilize the concept of transfer learning for electricity price forecasting by using data from four different markets. Our major novelties are listed below:

\begin{enumerate}
    \item We investigate various ways to combine data from different electricity markets when training neural networks.
    \item We propose a transfer learning scheme to leverage different market data when training deep neural networks (DNN) for the task of price prediction. 
    \item We analyze the usage of different exogenous variables for transfer learning. 
    \item We compare our transfer learning approach with open-access state-of-the-art DNN models and showcase the improved forecasting accuracy.
\end{enumerate}

The remainder of this paper is organised as follows. In Section \ref{sec:Related}, we first present an overview of the relevant literature in electricity price forecasting. Then, we review the literature on neural networks, local vs. global models, and transfer learning. In Section \ref{sec:materials}, we provide details of our data set and the pre-processing steps. In Section \ref{sec:methods}, we describe the methods used in this paper, which are mainly neural networks and transfer learning. Results and comparisons with state-of-the-art open-access models are presented in Section \ref{sec:experiments_results} and \ref{sec:Exogenous}, respectively; while Section \ref{sec:discussion_conclusion} discusses the findings of this paper in the context of the literature and proposes potential future work directions.


\section{Related Works}
\label{sec:Related}

In this section, we provide an overview of the relevant literature on electricity price forecasting, neural networks, local vs. global models, and transfer learning. 

\subsection{Electricity Price Forecasting}
\label{sec:imq}

Electricity price forecasting is a challenging task due to the nature of electricity prices. The seasonality in various frequencies, jumps to both sides and high volatility are the most challenging features of electricity prices. This tough task attracts academicians as well as practitioners from different fields. Therefore, electricity price forecasting is developed mainly in five branches: multi-agent, fundamental, reduced-form, statistical, and computational intelligence models \cite{weron2014}.

One important problem in electricity price forecasting is the non-generalizability of the results due to the unique structures of the markets. Moreover, applications are limited to single or mostly a few markets in most of the research. One important exception is Ziel and Weron's wide application in 12 different markets \cite{ziel2018day}. Their paper applies the statistical methods on various European markets and the GEFCom 2014 data \cite{hong2016probabilistic}. Furthermore, in a very recent paper, Lago et al. \cite{lago2021forecasting} proposes a review for electricity price forecasting by also comparing the performance of two open-source benchmark models (deep neural network (DNN) and Lasso models) in Belgium (BE), France (FR), Germany (DE), Nord Pool (NP), and U.S. PJM markets. From our point of view, forecasting electricity prices in such main markets give the opportunity to reach more generalized and robust results. We also follow this framework in our research and use all the covered European markets to forecast electricity prices.

\subsection{Neural Networks}

Recently, because of the statistical methods' limitations, various neural network models are applied to the electricity price forecasting problem\cite{hong2020locational}. A wide range study by Lago et al. shows that the machine learning methods outperform the statistical models \cite{lago2018forecasting}. In Lago et al. \cite{lago2018forecasting}, deep learning models such as convolutional neural networks (CNN), deep neural networks (DNN), standard recurrent neural networks (RNN), and long-short term memory (LSTM) are compared with the statistical models. In a similar fashion, according to Ugurlu et al. \cite{ugurlu2018}, the machine learning models outperform the statistical models. 
Moreover, the authors state that deep neural networks outperform single layer neural networks. Kuo and Huang \cite{kuo2018electricity} propose a deep neural network model, which combines CNN and LSTM to forecast electricity prices. The model is superior to various machine learning methods.

\subsection{Local vs. Global Models}
\label{sec:localglobal}

The question of how to use the time series data for forecasting model training has been investigated primarily on statistical models. The comparison of local (multiple model training for multiple time series) and global models (single model training for multiple time series) has been performed to understand the complexity vs. over-fitting phenomena in time series problems \cite{montero2021principles}. Recent papers \cite{salinas2020deepar, bandara2020forecasting} suggest that global models outperform the local models in terms of forecast accuracy in the application of neural network models. Our work follows a global approach, which can be defined as a stage-wise learning scheme. The global models aim to fit a single model to the data using multiple time series together at a single training scheme. Our fundamental difference with the prior works is the stage-wise training of the model, where the target task training is done last to ensure better performance.

\subsection{Transfer Learning}
\label{sec:transferlearning}

Research on transfer learning, which can be defined as transferring the information learned from a source domain to a similar target domain, has attracted attention from forecasting literature \cite{pan2009survey}. Collecting data is expensive, which makes transfer learning a viable option in a variety of forecasting applications. Tian et al. \cite{tian2019similarity} propose a neural-network-based smart meter forecasting scheme by using transfer learning. The main contribution of their model is that it needs less computational time compared to the traditional machine learning models. In a similar research \cite{laptev2018reconstruction}, authors apply a deep LSTM model by using transfer learning to forecast the residential scale electricity loads. Both works have relatively big data in the transfer learning applications. On the other hand, Hooshmand and Sharma \cite{hooshmand2019energy} apply transfer learning in their CNN model to forecast electricity demand, but with limited data. Their model outperforms the SARIMA model as well as simple CNN models. In a similar effort, Laptev et al. \cite{laptev2018reconstruction} show that transfer learning can be applied by RNNs with considerable success. Another electricity load forecasting paper \cite{xu2020hybrid} finds out that using another location's load as an additional source market improves the forecast accuracy up to 30\%. In related research, Zhou et al. \cite{zhou2020transfer} use the solar irradiance data to transfer the weights learned from the Long-Short Term Memory (LSTM) model to the photovoltaic power forecasting problem. According to their results, using transfer learning, especially with limited data, has positive effect on the prediction accuracy.

\subsubsection{Market Integration}
\label{sec:marketintegration}

The most related works in the electricity price forecasting literature for combining data from multiple markets were the market integration papers until very recent publications. The pionering work of Ziel et al. \cite{ziel2015forecasting} use the earlier announced Austrian electricity prices to forecast the German electricity prices. Lago et al. \cite{lago2018integrate} propose to integrate the French prices to forecast the Belgian prices. Chen et al. \cite{chen2019brim} apply a bidirectional integrated market-based LSTM model to forecast the French electricity prices. Their integration part follows the framework of Ziel et al. \cite{ziel2015forecasting} and the forecasts are compared with the benchmark models \cite{ziel2015forecasting, lago2018integrate, lago2018forecasting}.

\subsubsection{Transfer Learning in Electricity Price Forecasting}
\label{sec:tlepf}

There are mainly two papers \cite{luo2019two, yang2021real} published very recently about the applications of transfer learning in electricity prices. Luo and Weng \cite{luo2019two} use two-stage supervised learning, which can be called a version of transfer learning. They use  wind power generation as the first stage and historical electricity prices as the second stage to forecast the electricity prices. One of the most important findings in this research is that using the best training interval for different data sources has a positive effect. Moreover, they also conclude that using two-stage learning decreases forecast errors. In a very recent paper, Yang and Schell \cite{yang2021real} forecast the electricity prices for wind farms. They find out that using the GRU model with transfer learning outperforms all the benchmarks in terms of forecast accuracy. 

Our paper differs from these papers as it uses various data usage methods to compare the performance of transfer learning. Additionally, our paper is the first paper which that uses various day-ahead markets' data as the source market to forecast the electricity prices for the target day-ahead market.

\section{Data}
\label{sec:materials}

\begin{figure}
\begin{center}
    \includegraphics[width=\linewidth]{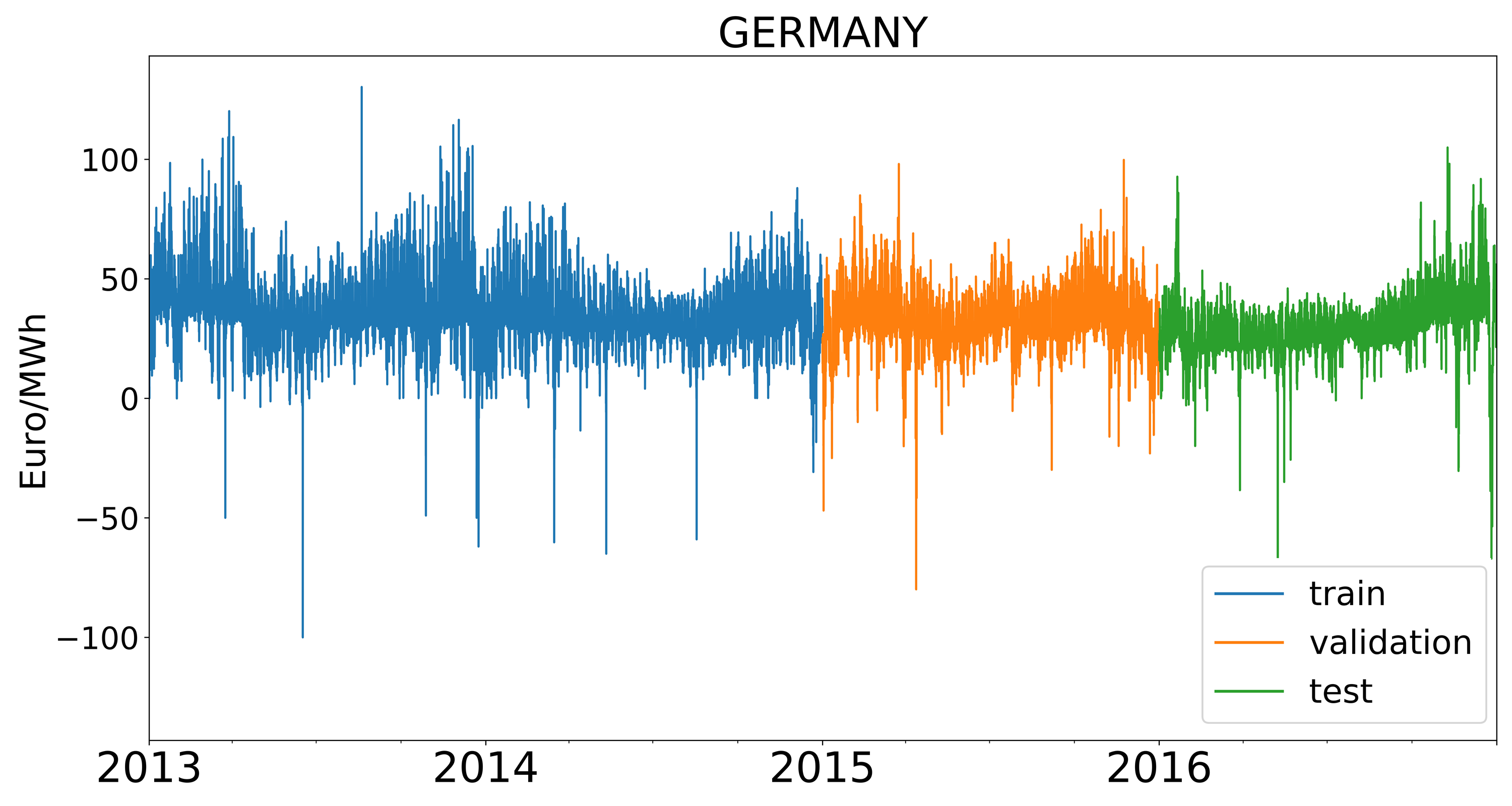}
    \caption{Training, validation and test parts for the target data (In the example of German data).}
    \label{fig:germany}
\end{center}
\end{figure}

In this paper, four different day-ahead electricity markets (BE, DE, FR, NP) are examined. The main experiment set up is an ablation study. We use a fixed model and a common exogenous variable (temperature) for four markets. In this way, we can clearly analyze the contribution of transfer learning. The data for these markets are obtained in hourly frequency \cite{Entsoe, Opsd, Opsd2021} and forecasts are done for each hour of the following day. Hourly temperature data is the population-weighted mean across all NASA MERRA-2 grid cells within the given country \cite{Opsd2021}. The training period is from 01.01.2013 to 31.12.2014, validation period is from 01.01.2015 to 31.12.2015 and test period is from 01.01.2016 to 31.12.2016. Due to daylight saving time change, for missing hours, average of the previous hour and the following hour is used. When there are two prices for the same hour due to the same effect, the average of these prices is utilized. Figure \ref{fig:germany} demonstrates the training, validation, and test parts of the data in the German market example. The forecasts are performed tests for days of the entire year to diminish the seasonality effect on average test results. In addition to the lagged electricity prices as the endogenous variable, we also used lagged temperatures as the exogenous variable and the days of the week as the dummy variable. 

\subsection{Descriptive Statistics}

\begin{figure}
\centering
\begin{subfigure}[b]{.49\linewidth}
\includegraphics[width=\linewidth]{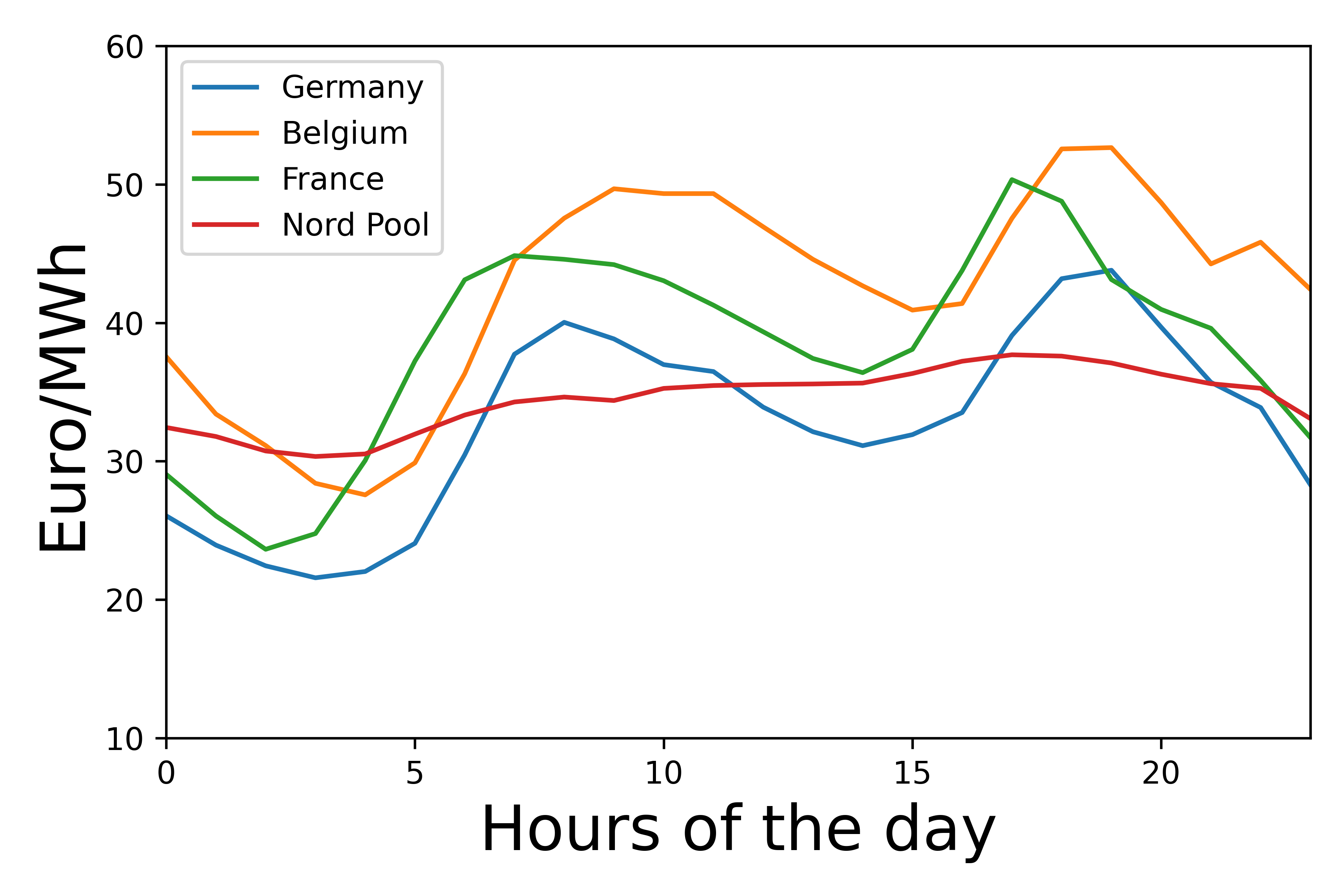}
\caption{}\label{fig:daily}
\end{subfigure}
\begin{subfigure}[b]{.49\linewidth}
\includegraphics[width=\linewidth]{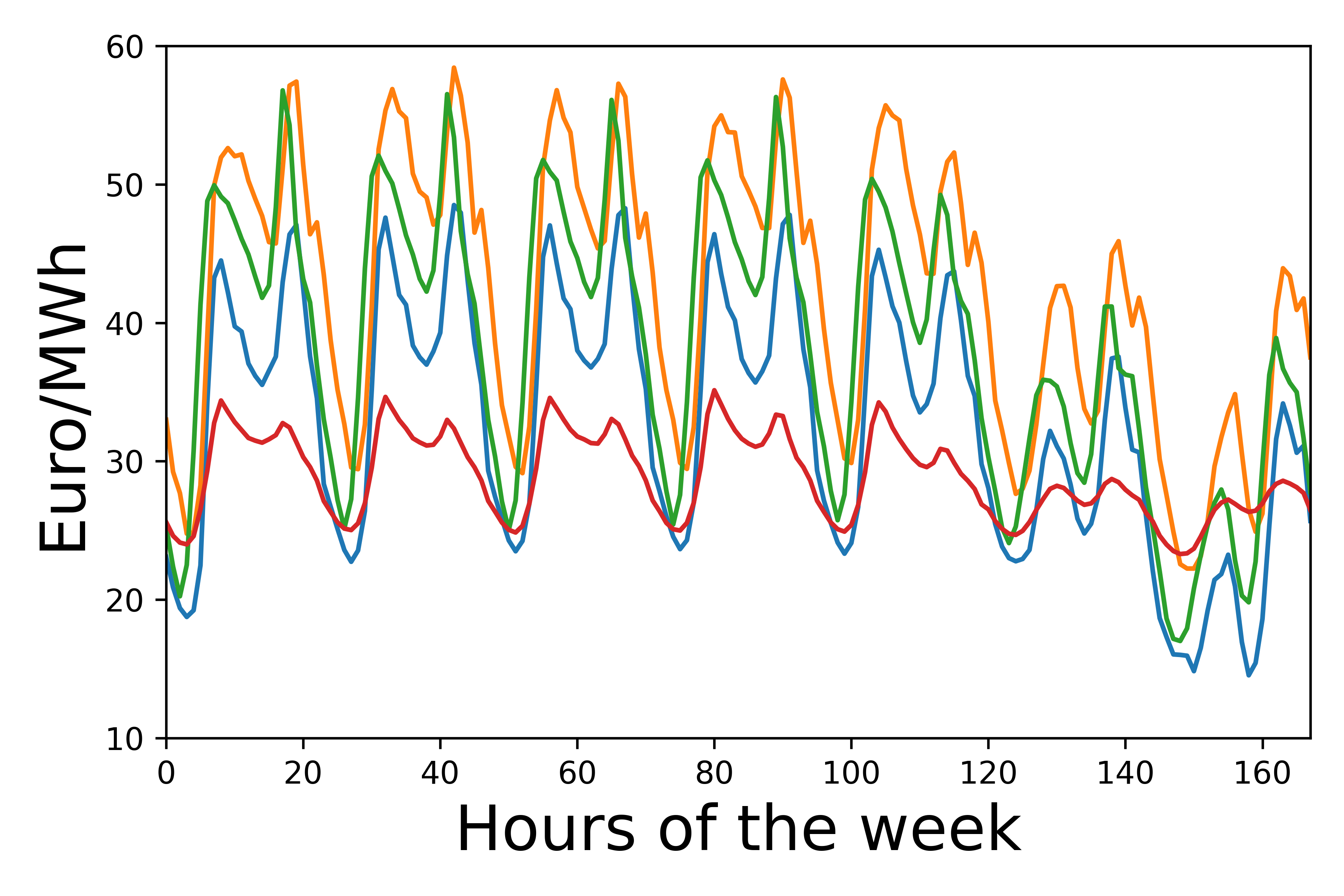}
\caption{}\label{fig:weekly}
\end{subfigure}

\caption{(a) Average electricity prices for all markets according to the hours of the day, (b) Average electricity prices for all markets according to the hours of the week.}
\label{fig:mean}
\end{figure}

In Figure \ref{fig:daily}, we illustrate German, French and Belgian markets co-movement in terms of 24-hour averages from 2013 to 2016. On the other hand, The NP market behaves different than the other markets which might be due to the large share of dispatchable and flexible hydropower. In Figure \ref{fig:weekly}, the same pattern can be observed. In this graph, 168-hour averages from 2013 to 2016 are given for all the countries. Both, low price and volatility levels of the NP market are the most striking outcomes of the graph.

In Table \ref{table:markets}, we showcase general price levels are in a decreasing trend from 2013 to 2016 in all evaluated countries. For instance, in DE, high share of wind energy generation decreases electricity price levels for the country. Furthermore, in NP, standard deviations are low compared to other countries because of the smoother generation regarding large share of hydropower.

\begin{table}[htb]
\centering
\caption{Mean ± standard deviation values of all markets' data from 2013 to 2016}
\footnotesize 
\begin{tabular}{ccccccl}
        \hline
                     & Belgium              & Germany               & France               & Nord Pool                          &  \\\hline
2013                 & 47.45 ± 19.24        & 37.78 ± 16.46        & 43.24 ± 20.26        & 38.10 ± 6.94          \\
2014                 & 40.79 ± 12.67        & 32.76 ± 12.77        & 34.63 ± 13.90        & 29.60 ± 5.34           \\
2015                 & 44.50 ± 18.87        & 31.62 ± 12.66        & 38.47 ± 12.94        & 20.97 ± 7.91          \\
2016                 & 36.46 ± 20.73        & 28.97 ± 12.48        & 37.50 ± 16.53        & 26.91 ± 8.95          \\\hline
\end{tabular}
\label{table:markets}
\end{table}

\subsection{Pre-processing of the Data}

Our data set includes prices, hourly temperatures, and the day of the week dummy variables. We propose to generate data samples for multi-step forecasting with 343 inputs and 24 targets as shown in Figure \ref{fig:DNN}. Previous 168 hours' prices, previous 168 temperature values, and 7 dummy variables representing day of the week are the input features. If we assume that we are at day d and time h, the next 24 prices are the target of the model. Training samples are constituted by the hourly rolling window and the problem becomes a conventional supervised learning problem. Every training sample is independent, which means that we can shuffle and change the order of the data. By using this method, we can create the samples; stack selected useful market data and shuffle the samples.
We can use the stacking approach for two purposes. The first one is using source markets' data for pre-training and the second one is using stacked data for multi-task learning as shown in Figure \ref{fig:comp}.

\section{Methods}
\label{sec:methods}


In this section, we explain the deep neural network (DNN) model that we utilized for electricity price forecasting. We give brief information about naive and linear models that we use as benchmarks. We also introduce the concept of transfer learning and the implementation details.

\subsection{Data Transformation}
\label{sec:normalize}
Variance stabilizing transformation (VST) is an important tool to smooth the effects of the spikes in electricity prices \cite{uniejewski2017variance}. Firstly, we normalize the data with Median-Mad transformation defined as:

\begin{equation}
  S_{d,h} =  (P_{d,h} - Median( P_{d,h})) / MAD( P_{d,h}) 
  \label{eq:normalize}
\end{equation}

where $P_{d,h}$ is unnormalized data at day d and hour h, MAD is \emph{median absolute deviation} and $S_{d,h}$ is normalized data. After normalization we apply \emph{area hyperbolic sine(asinh)} \cite{uniejewski2017variance}:

\begin{equation}
  Y_{d,h} = log(S_{d,h} +  \sqrt{S_{d,h}^2+1})
  \label{eq:asinh}
\end{equation}
where $Y_{d,h}$ represents the transformed data at day d and hour h. The inverse of asinh transformation is \emph{hyperbolic sinus(sinh)}. We apply inverse transformation to the output of the model $\hat{Y}_{d,h}$ in reverse order to reach point forecasts $\hat{P}_{d,h}$.

\subsection{Basic Deep Neural Network Model for Electricity Price Forecasting}
\begin{figure}[H]
\centering
    \includegraphics[width=\linewidth*3/4]{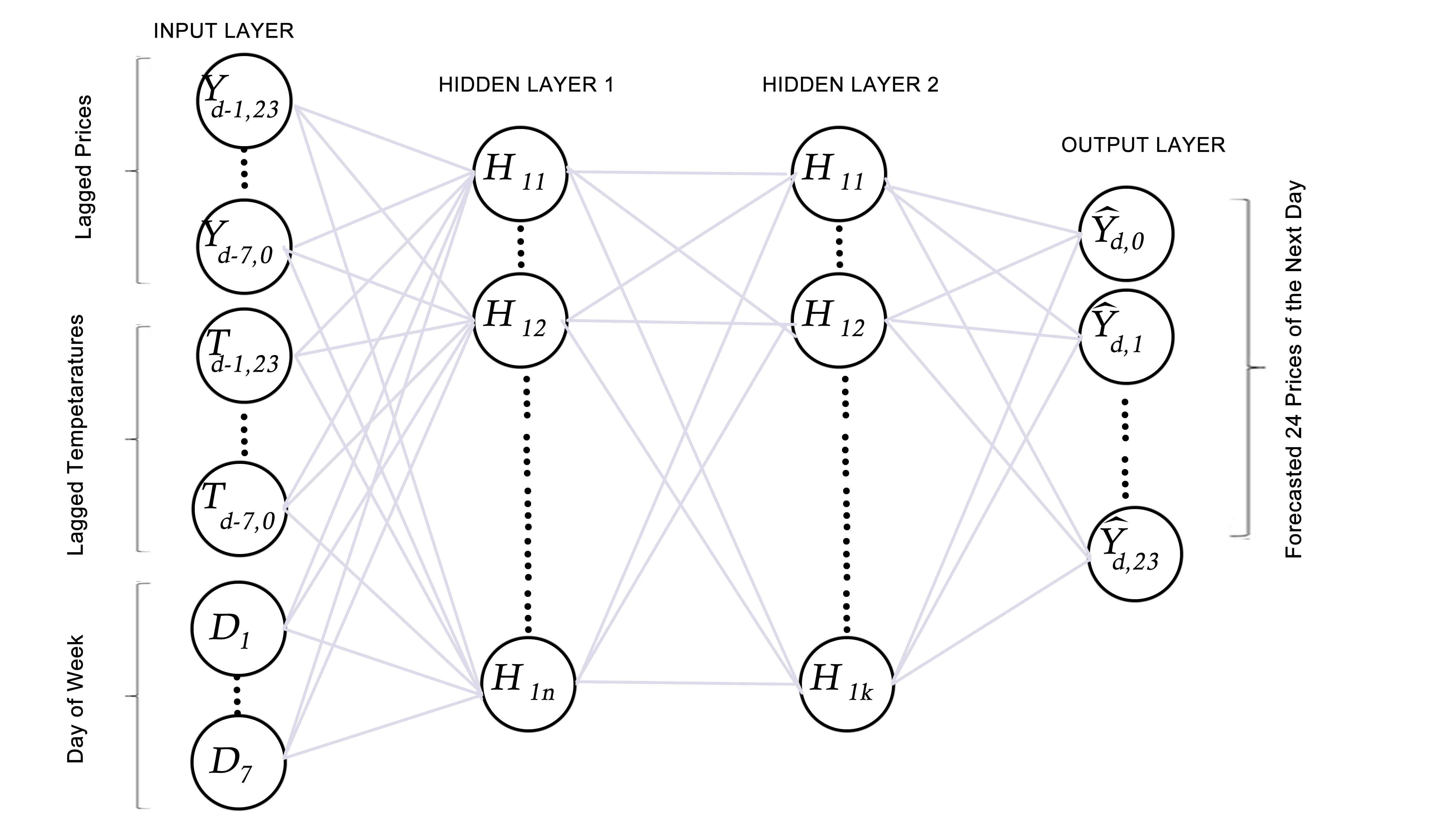}
    \caption{DNN model with two hidden layers, 343 input and 24 output neurons.}
    \label{fig:DNN}
\end{figure}

We utilize a DNN model to forecast the next 24 hours electricity prices. DNN model is constructed with 2 hidden layers and an output layer which contains 24 outputs as illustrated in Figure \ref{fig:DNN}. The input of the model is defined as: 

\begin{equation}
X = [Y_{d-1,23}, Y_{d-1,22}, ..., Y_{d-7,0}, T_{d-1,23},T_{d-1,22},...,T_{d-7,0}, D_1, ..,D_7 ]^T 
  \label{eq:dnn_in}
\end{equation}

where $Y_{d,h}$'s are lagged prices on day d and hour h, $T_{d,h}$'s are lagged temperatures and $D$'s are dummy representations of the day of week. The output of the model is given below:

\begin{equation}
Y = [Y_{d,0}, Y_{d,1}, ..., Y_{d,23}]^T 
  \label{eq:dnn_out}
\end{equation}

where $Y$ is a vector of hourly electricity prices of the next 24 hours. Details of the model are illustrated in Figure \ref{fig:DNN}, where $H_{1n}$ is the number of neurons in hidden layer 1 and $H_{2k}$ is the number of neurons in hidden layer 2.  The basic model is trained with only single market's data.

\subsection{Integrate}
We construct the integrate model by combining several markets' inputs and feed them to a single DNN model. The inputs of the model is defined as:

\begin{equation}
X = [M_1,M_2, .. , M_n, D_1, ..,D_7 ]^T 
  \label{eq:int_in}
\end{equation}

where $M$ represents each market's temperature and price values in equation \ref{eq:dnn_in}. $n$ is the number of integrated markets. We add dummies at the end of the inputs. The output of the model is forecasts for the target market:

\begin{equation}
Y_{target} = [Y_{d,0}, Y_{d,1}, ..., Y_{d,23}]^T 
  \label{eq:int_out}
\end{equation}

\subsection{Pretrain-Finetune}
\label{sec:TL_method}
\begin{figure}[H]
\centering
    \includegraphics[width=\linewidth*3/4]{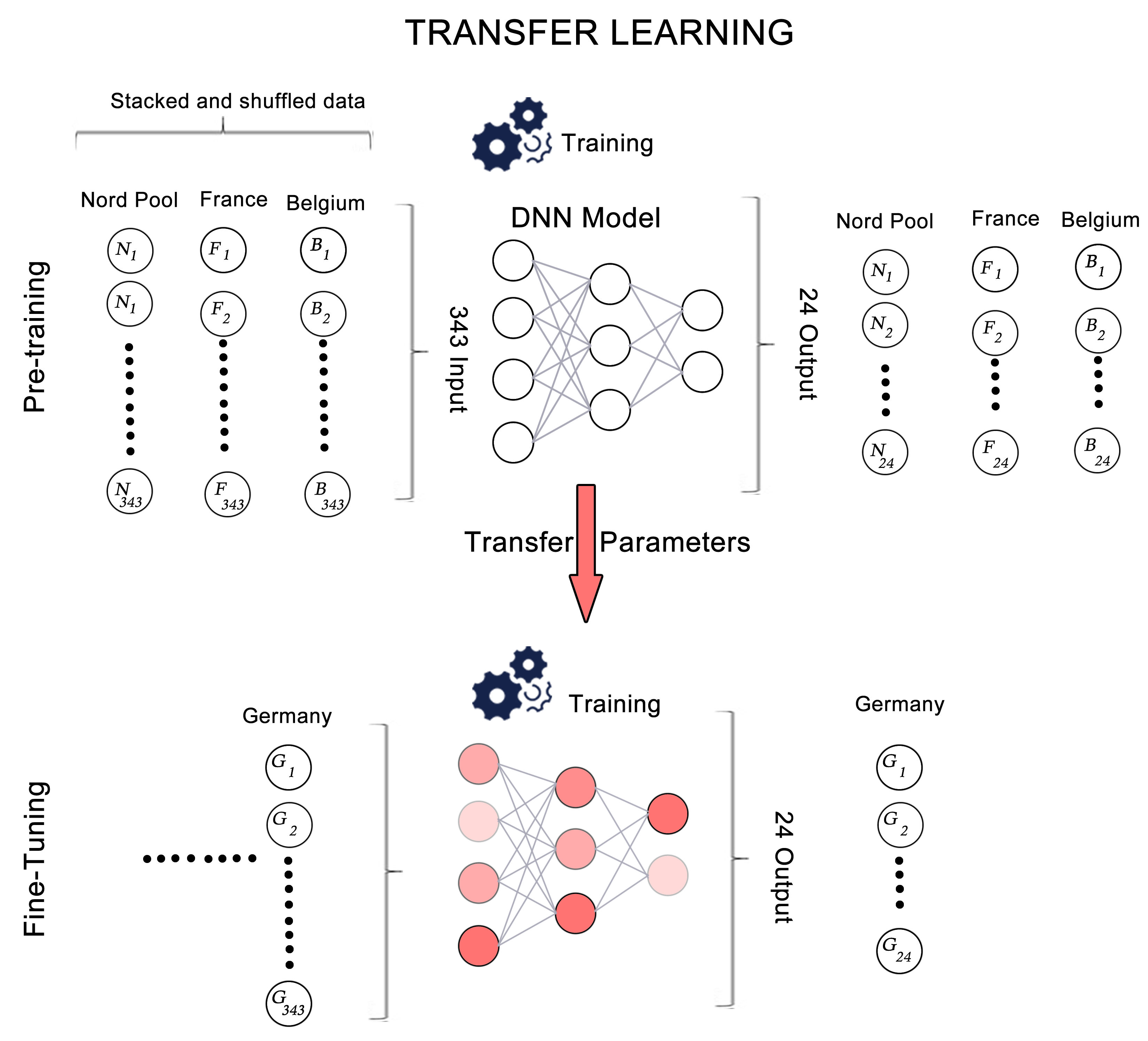}
    \caption{Transfer learning scheme for the German Market. Data from source markets (NP, FR, and BE) is used for pre-training the model. The pre-trained model parameters are utilized to initialize the fine-tuning step (re-training on the German market).}
    \label{fig:transfer}
\end{figure}

Given a source domain $D_S$ and learning task $L_S$, a target domain $D_T$ and learning task $L_T$, transfer learning aims to improve the learning of the target predictive function $f_T$(·) in $D_T$ by using the knowledge in $D_S$ and $L_S$, where $D_S \neq D_T$, or $ L_{S} \neq L_{T} $ \cite{pan2009survey}. Our fundamental idea is to use a pre-trained model to warm start the training of the DNN model for the target domain. Figure \ref{fig:transfer} illustrates the example for the German electricity market, where we did a pre-training by using data from NP, FR, and BE. The final parameter setup of the pre-trained model is used as a starting point for the fine-tuning step (re-training on the German market). We repeat this process for all four markets and report the prediction accuracy in the results section using a different number of markets to warm start in each setup.

\subsection{Multi-task Learning}
\label{sec:multi-task}
\begin{figure}[H]
\centering
    \includegraphics[width=\linewidth]{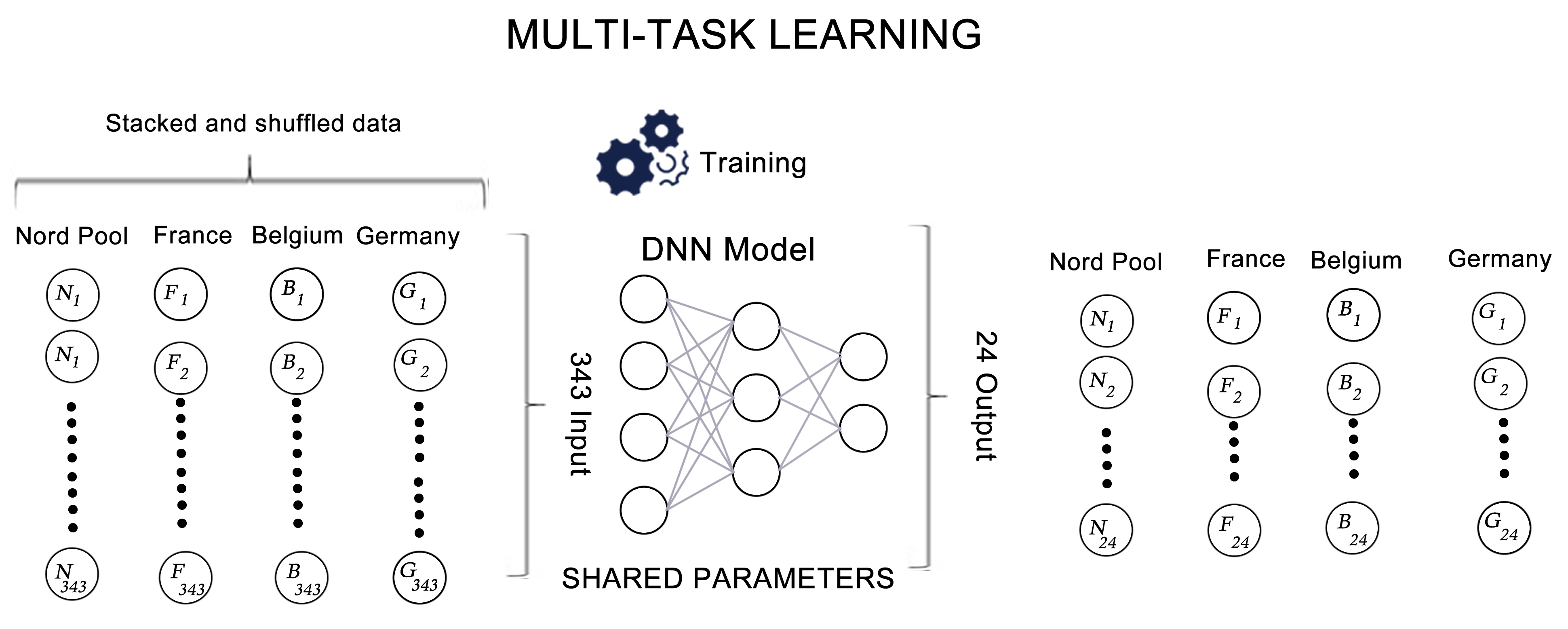}
    \caption{Multi-task learning scheme. Four markets are trained simultaneously with a single model in a single stage training strategy. The trained model is used for all markets at the test time.}
    \label{fig:multi_task}
\end{figure}

Multi-task learning (MTL) is a kind of inductive transfer learning method that helps generalization. MTL differs from common transfer learning, related tasks are learned simultaneously \cite{caruana1997multitask, pan2009survey}. Sharing parameters may increase generalization, even if we optimize only one loss function. Moreover, MTL also provides implicit data augmentation that helps generalization. Figure \ref{fig:multi_task} illustrates the training strategy for the multi-task model. We stack data from several markets and feed into our model in Figure \ref{fig:DNN}. Model's validation data consists of stacked validation data from multiple markets and there is only single-stage training for a single model, unlike the transfer learning setup we introduced in Section \ref{sec:TL_method}.  


\subsection{Pre-train}

This method is making predictions by the pre-trained model without fine-tuning with target market's data. Although this method is not thought as a forecasting method, it gives meaningful information about the distance between source and target tasks. We aim to analyze the influence of fine-tuning with experiments of the pre-trained model.

\subsection{Implementation Details}
\label{sec:implementationdetails}
Implementations and experiments are done by using the Python programming language and its' libraries (e.g. Tensorflow, Keras, and the scikit-learn  \cite{pedregosa2011scikit}). The training of the DNN model consists of two fundamental components: a loss function and an optimization algorithm that minimizes the loss function. We set the number of neurons for the first hidden layer $H_{1n}$ equal to 64 and the number of neurons for the second hidden layer $H_{ 2k}$ equal to 32 empirically. We use Adam optimiser with a learning rate 0.001 to minimize the MAE loss function defined in equation \ref{eq:eq_mae}. The training finishes by early stopping, when the network does not significantly improve its performance on the validation set after 10 epochs patience (maximum of 1000 epochs allowed). We set the batch size to 64, which we optimized empirically. Data of 2015 is used as the validation data throughout the study. In the fine-tuning process, we set all layers untrainable except the output layer and we update the learning rate as 0.0001. The patience of early stopping for fine-tuning is equal to 1 epoch. The transfer learning model used in the final prediction is selected utilizing a different set of markets for pre-training in each case according to the validation set.

\subsection{Evaluation Metrics}
\label{sec:metrics}
Similar to Lago et al. \cite{lago2021forecasting}, we apply widely used evaluation metrics in electricity price forecasting literature. Scale-dependent metrics \emph{mean absolute error}(MAE) and \emph{root mean square error}(RMSE) are defined as: 

\begin{equation}
  MAE = \frac{1}{24 N_d} \sum_{d=1}^{ N_d} \sum_{h=1}^{24}|P_{d,h} - \hat{P}_{d,h}|
  \label{eq:eq_mae}
\end{equation}

\begin{equation}
  RMSE = \sqrt{\frac{1}{24 N_d} \sum_{d=1}^{ N_d} \sum_{h=1}^{24}(P_{d,h} - \hat{P}_{d,h}})^2
  \label{eq:eq_rmse}
\end{equation}

where $P_{d,h}$ is the actual and $\hat{P}_{d,h}$ is the predicted price at $h$’th hour of the day $d$. $N_d$ is number of days. Since these metrics do not help much with the comparison of the values between different markets or different time intervals, \emph{mean absolute percentage error} (MAPE) which is based on percentage errors is a viable option \cite{hyndman2006another}. On the other hand, MAPE cannot be calculated if there is an actual value of zero in the series and it also has bigger penalty on positive errors. To avoid these problems, we utilize a modified version of MAPE called \emph{symmetric mean absolute percentage error} (sMAPE). Another measure \emph{relative mean absolute error} (rMAE) provides evaluation on different datasets. sMAPE and rMAE are defined as:

\begin{equation}
  sMAPE = \frac{1}{24 N_d} \sum_{d=1}^{ N_d} \sum_{h=1}^{24} 2 \frac{|P_{d,h} - \hat{P}_{d,h}|}{|P_{d,h}| + |\hat{P}_{d,h}|}
  \label{eq:eq_rmae}
\end{equation}

\begin{equation}
  rMAE = \frac{\frac{1}{24 N_d} \sum_{d=1}^{ N_d} \sum_{h=1}^{24}|P_{d,h} - \hat{P}_{d,h}|}{\frac{1}{24 N_d} \sum_{d=1}^{ N_d} \sum_{h=1}^{24}|P_{d,h} - \hat{P}^{naive}_{d,h}|}
  \label{eq:eq_smape}
\end{equation}

$\hat{P}^{naive}_{d,h}$ is the electricity price forecasted by naive model which is explained in equation \ref{eqn:naive}.

\subsection{The Naive Benchmark}
\label{sec:naive}
We use the naive model as the first benchmark which also enables the calculation of rMAE metric for the evaluation. The naive model \cite{nogales2002forecasting} is defined as:

\begin{equation}
\centering
   \hat{P}_{d,h}= 
    \begin{cases}
        P_{d-7,h},& \text{Monday, Saturday, Sunday } \\
        P_{d-1,h},              & \text{Tuesday, Wednesday, Thursday, Friday}
        \label{eqn:naive}
    \end{cases}
\end{equation}

where $ \hat{P}_{d,h}$ is the forecasted electricity price at h'th hour of the day d.

\subsection{The Linear Benchmark}

We use a linear model with lasso regularization, Lasso Estimated Auto-Regressive (LEAR) from \cite{lago2021forecasting}. This model is structured like a fARX model which utilizes L1 regularization \cite{uniejewski2016automated}. We utilize 24 models for each hour of the day. The compact representation of the fARX model is as follows: 

\begin{equation}
  \hat{Y}_{d,h} =  \sum_{i=1}^{n}  \beta_{h,i} Y_{d,h,i} + \epsilon_{d,h}
  \label{eq:eq_farx}
\end{equation}

where $n$ = 343 regressors so that the input of the model is the same input vector of the DNN model and $\beta_{h,i}$'s are their coefficients. The lasso method provides shrinkage and reduces variance:

\begin{equation}
  \hat{\beta} =  \arg\min_{\beta_{h,i}} \{ RSS + \lambda \sum_{i=1}^{n} |\beta_{h,i} |\}
  \label{eq:eq_lasso}
\end{equation}

where $\lambda$ is penalty term of coefficients, which is set to 0.001 empirically in our experiments.
\section{Experiments and Results}
\label{sec:experiments_results}

In this section, we provide the experimental results for different usage of training data from four markets. Firstly, we show the influence of using exogenous variables in the transfer learning setup. Secondly, we provide quantitative results on different usages of data, where our transfer learning results are based on fine-tuning the pre-trained models. Thirdly, we do a thorough analysis of transfer learning on using various portions of data. Finally, we perform Diebold-Mariano \cite{Diebold1995} test for checking the statistical significance of the performance difference between the models.

\subsection{Analysis on Exogenous Variable}

In order to see the influence of exogenous variable on the final transfer learning framework, firstly, we trained the model only with 168-lagged price values from source markets. Then, we used transfer learning to fine-tune the model, which is named as \emph{without exogenous} in Table \ref{table:Exogenous}. We compare this method with the transfer learning method including exogenous variables (168 lagged prices, 168 lagged temperature values, 7 dummy variables for the days of the week), which is named as \emph{with exogenous}. The results indicate the superior performance of using exogenous variables in the context of transfer learning for all four markets. Therefore, we prefer to continue our study by using the \emph{with exogenous} model.

\begin{table}[H]
\centering
\caption{Transfer learning results for MAE with exogenous and without exogenous inputs.}
\begin{tabular}{lcccc}
Transfer          & Belgium              & France               & Germany              & Nord Pool            \\\hline
With Exogenous    & \cellcolor[HTML]{C0C0C0}\textbf{5.68}                & \cellcolor[HTML]{C0C0C0}\textbf{4.21}                & \cellcolor[HTML]{C0C0C0}\textbf{4.22}                & \cellcolor[HTML]{C0C0C0}\textbf{1.89}                 \\
Without Exogenous & 6.02                 & 4.29                 & 4.44                 & 1.97                 \\\hline
                  & \multicolumn{1}{l}{} & \multicolumn{1}{l}{} & \multicolumn{1}{l}{} & \multicolumn{1}{l}{}
\end{tabular}
\label{table:Exogenous}
\end{table}

\subsection{Quantitative Results}

\begin{figure}[ht]
\centering
    \includegraphics[width=\linewidth*2/4]{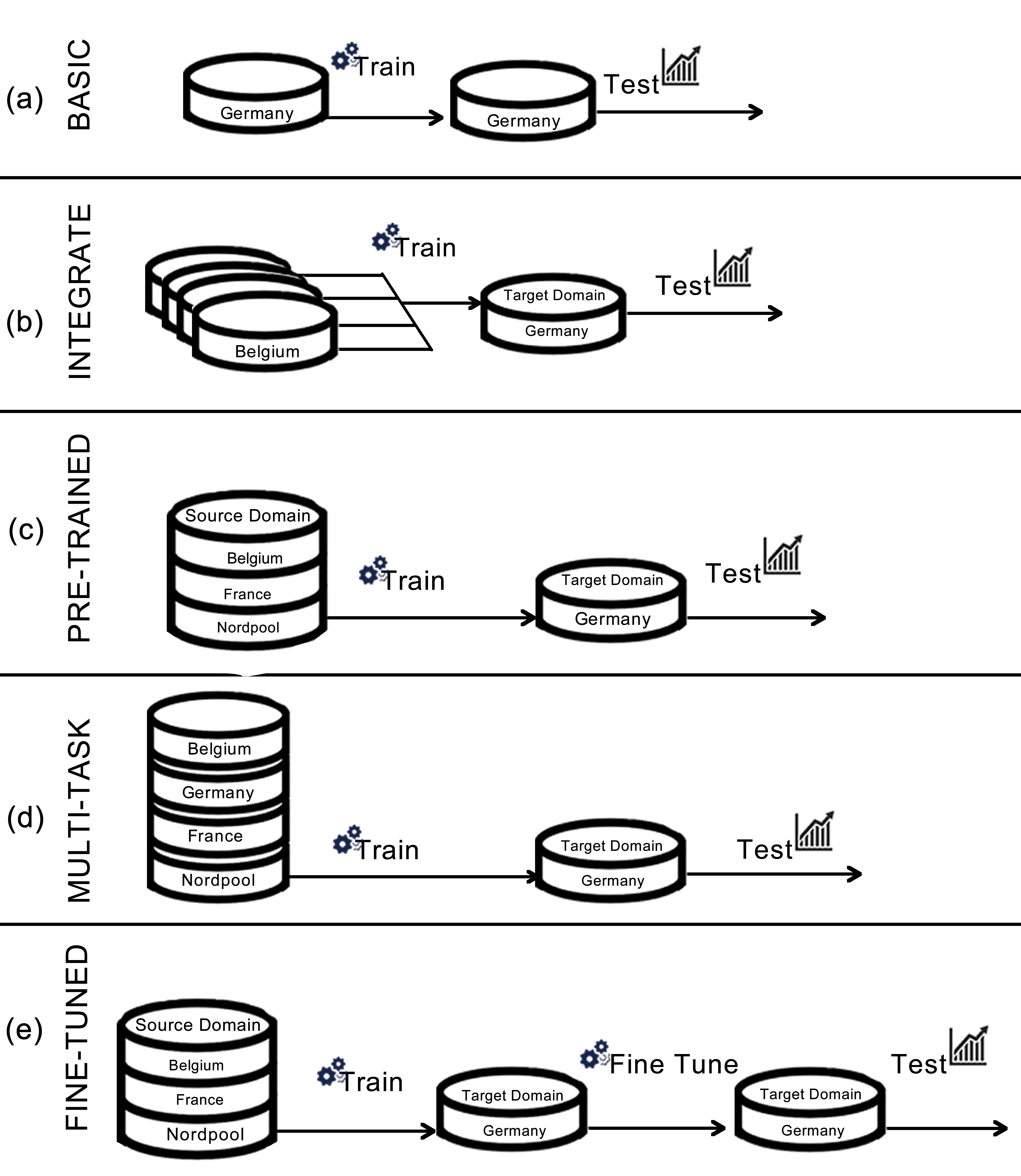}
    \caption{Comparison of the data usage methods with respect to the DE market test. In all scenarios except the Integrate, the DNN model in Figure \ref{fig:DNN} which has 343 inputs and 24 outputs is used. The Integrate model has n $\times$ 343 inputs and 24 outputs where n is input market count.}
    \label{fig:comp}
\end{figure}

We perform a wide quantitative analysis to see different variations of using training data and its influence on the final performance as depicted in Figure \ref{fig:comp}. We compare using a transfer learning framework Pretrain-Finetune), with the multi-task network for all available data from four different markets for single training (Multi-task), with using a basic training on a single market (Basic), a model trained on source markets and tested on the target one without re-training the network on its own data (Pre-trained), an integration model, which is similar to the application in \cite{lago2018integrate}, where all four markets data is integrated as input variables into a single model (Integrate). For all models of comparison, different combinations of markets are tested and best model results are reported. Table \ref{table:Quan} indicates that the transfer model gives the lowest errors in terms of MAE, RMSE, sMAPE and rMAE for FR, DE and NP markets. For BE market, while MAE and sMAPE values are the lowest for the transfer model, rMAE numbers are equally low for pre-trained and transfer models and RMSE value is the lowest for the pre-trained model. 


\begin{table}[htb]
\centering

\caption{Quantitative results on test set for MAE, RMSE, sMAPE and rMAE (tested with Naive method) metrics. The results indicate that transfer learning improves the performance in almost all scenarios.}
\begin{adjustbox}{width=1\textwidth}
\small
\begin{tabular}{ccccccccc}
\multicolumn{1}{l}{}       & Error  & Naive & LEAR & Basic & Integrate & Pre-trained    & Multi-Task & Fine-tuned       \\\hline
\multirow{4}{*}{Belgium}   & MAE   & 6.73 & 6.23  & 6.05  & 6.22      & 5.68           & 5.84       & \cellcolor[HTML]{C0C0C0}\textbf{5.66}  \\
                           & RMSE  & 13.07 &11.77 & 11.64 & 11.56     & \cellcolor[HTML]{C0C0C0}\textbf{10.97} & 11.11      & 10.99          \\
                           & sMAPE & 19.01 & 17.34 & 17.36 & 17.25     & 15.57          & 15.90      & \cellcolor[HTML]{C0C0C0}\textbf{15.50} \\
                           & rMAE  & 1.00 & 0.92    & 0.89  & 0.92      & \cellcolor[HTML]{C0C0C0}\textbf{0.84}  & 0.86       & \cellcolor[HTML]{C0C0C0}\textbf{0.84}  \\\hline
\multirow{4}{*}{France}    & MAE   & 5.64 & 4.59  & 4.45  & 4.63      & 4.46           & 4.66       & \cellcolor[HTML]{C0C0C0}\textbf{4.21}  \\
                           & RMSE  & 10.13 & 8.28 & 8.36  & 8.71      & 8.50           & 8.60       & \cellcolor[HTML]{C0C0C0}\textbf{7.92}  \\
                           & sMAPE & 17.40 & 13.68 & 13.16 & 13.45     & 12.87          & 13.61      & \cellcolor[HTML]{C0C0C0}\textbf{12.29} \\
                           & rMAE  & 1.00 & 0.81    & 0.78  & 0.81      & 0.79           & 0.82       & \cellcolor[HTML]{C0C0C0}\textbf{0.74}  \\\hline
\multirow{4}{*}{Germany}   & MAE   & 6.18 & 4.76 & 4.37  & 4.81      & 4.46           & 4.34       & \cellcolor[HTML]{C0C0C0}\textbf{4.22}  \\
                           & RMSE  & 10.56 & 7.51 & 6.97  & 7.51      & 7.25           & 7.03       & \cellcolor[HTML]{C0C0C0}\textbf{6.83}  \\
                           & sMAPE & 25.37 & 18.75 & 17.56 & 19.31     & 17.73          & 17.14      & \cellcolor[HTML]{C0C0C0}\textbf{16.88} \\
                           & rMAE  & 1.00  &0.77   & 0.70  & 0.77      & 0.72           & 0.70       & \cellcolor[HTML]{C0C0C0}\textbf{0.68}  \\\hline
\multirow{4}{*}{Nord Pool} & MAE   & 2.52 & 1.96  & 2.06  & 2.19      & 2.76           & 2.11       & \cellcolor[HTML]{C0C0C0}\textbf{1.89}  \\
                           & RMSE  & 6.01 & 4.79 & 4.86  & 5.07      & 5.39           & 4.76       & \cellcolor[HTML]{C0C0C0}\textbf{4.51}  \\
                           & sMAPE & 8.83 & \cellcolor[HTML]{C0C0C0}\textbf{6.71}  & 7.39  & 8.35      & 10.44          & 7.41  & 6.93  \\
                           & rMAE  & 1.00 & 0.77    & 0.81  & 0.86      & 1.09           & 0.81       & \cellcolor[HTML]{C0C0C0}\textbf{0.75} \\\hline
\end{tabular}
\end{adjustbox}
\label{table:Quan}
\end{table}

\subsection{Analysis on amount of training data}
\label{sec:Transfer_result}

\begin{figure}[ht]
\begin{centering}
\begin{tabular}{cc}
\includegraphics[width=6cm]{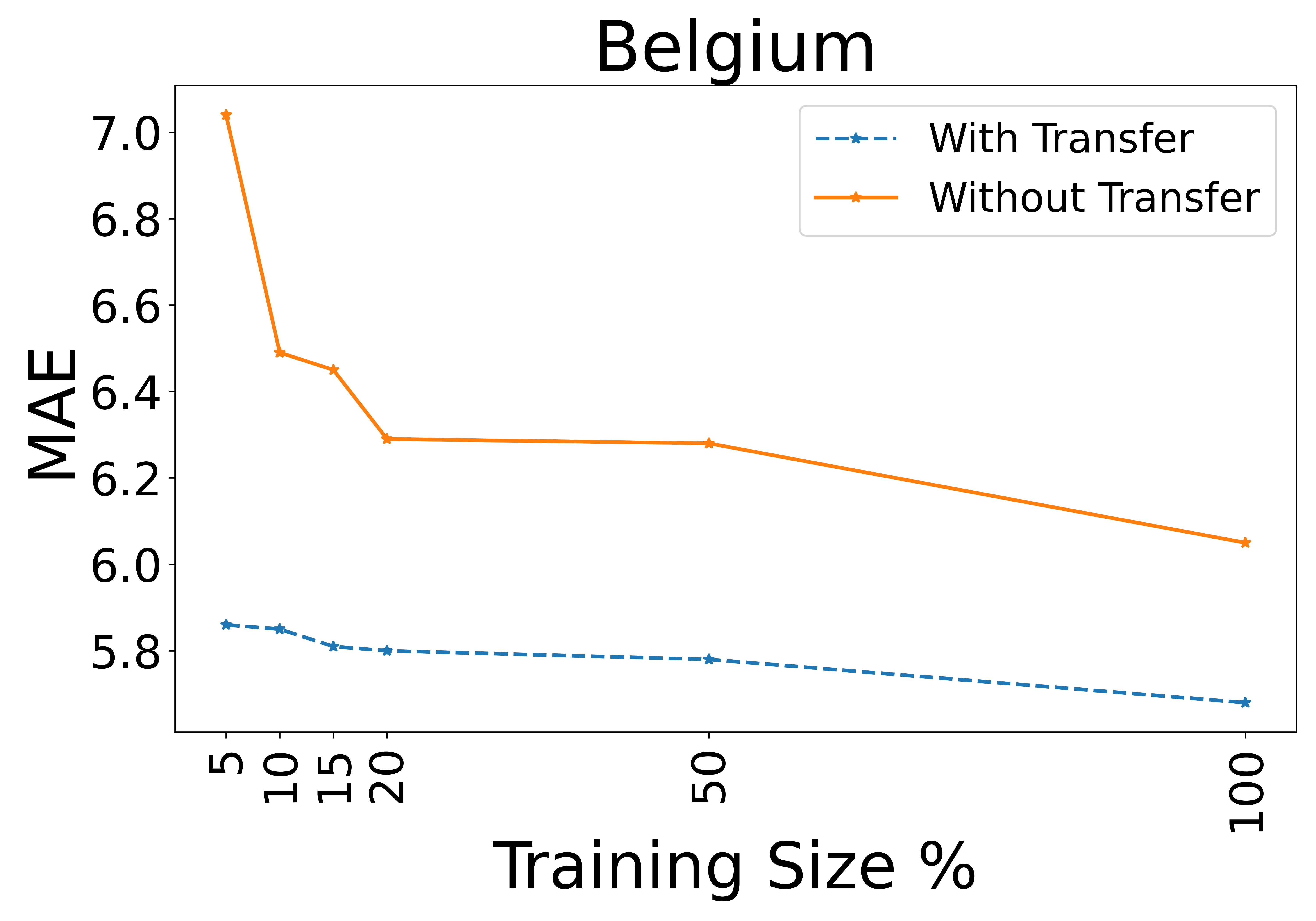} &  
\includegraphics[width=6cm]{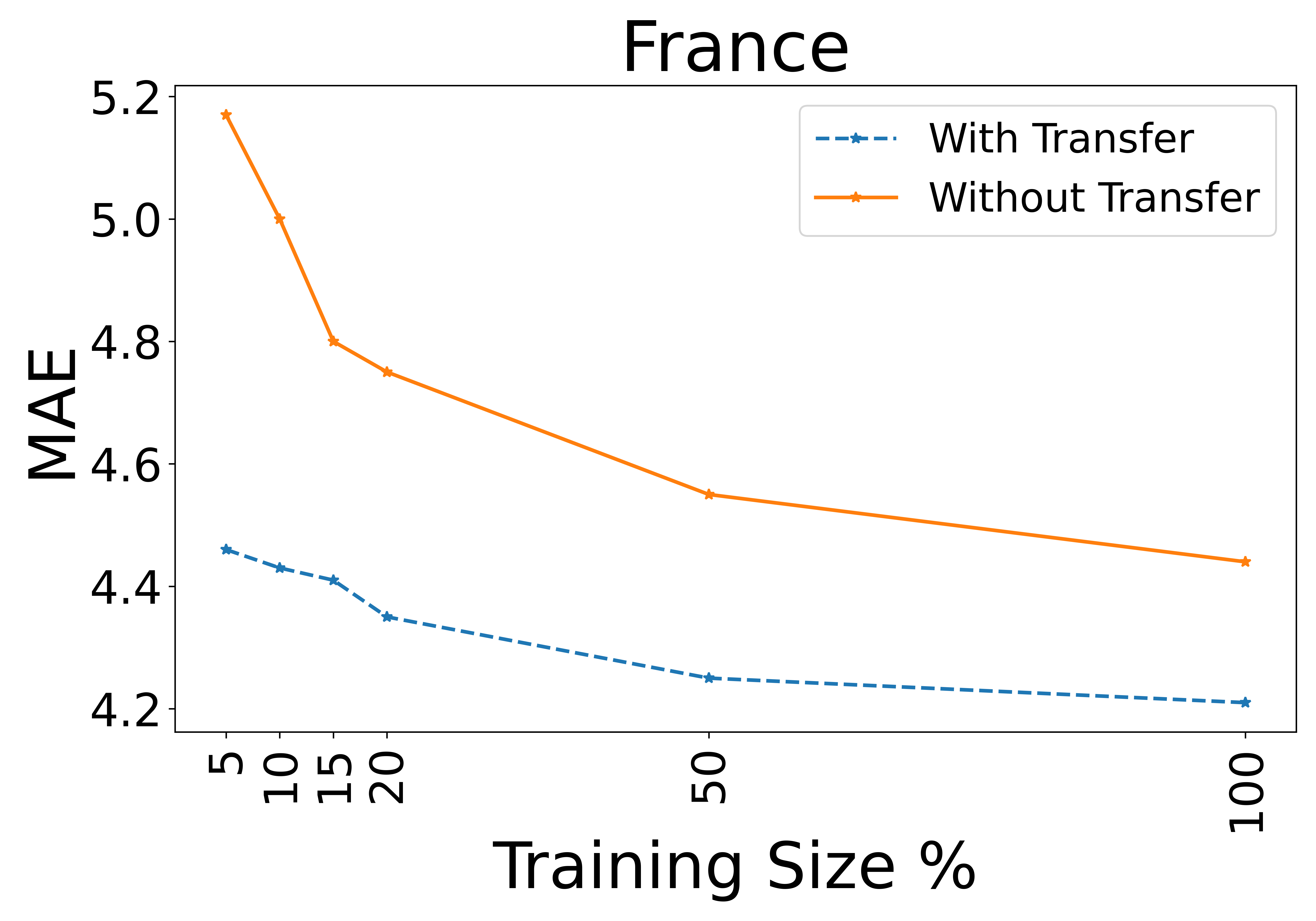} 
\end{tabular}
\begin{tabular}{cc}
 \includegraphics[width=6cm]{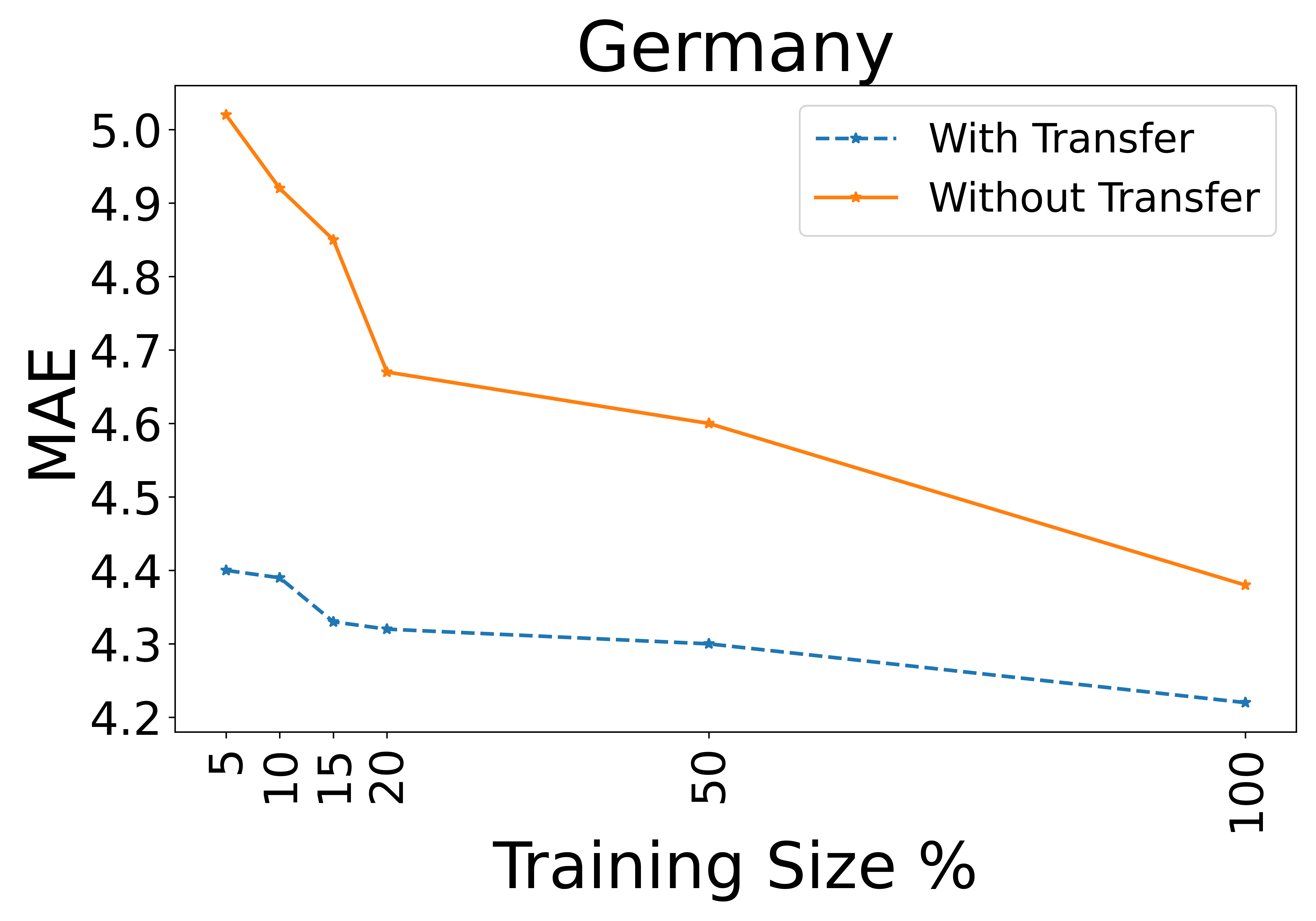} & 
 \includegraphics[width=6cm]{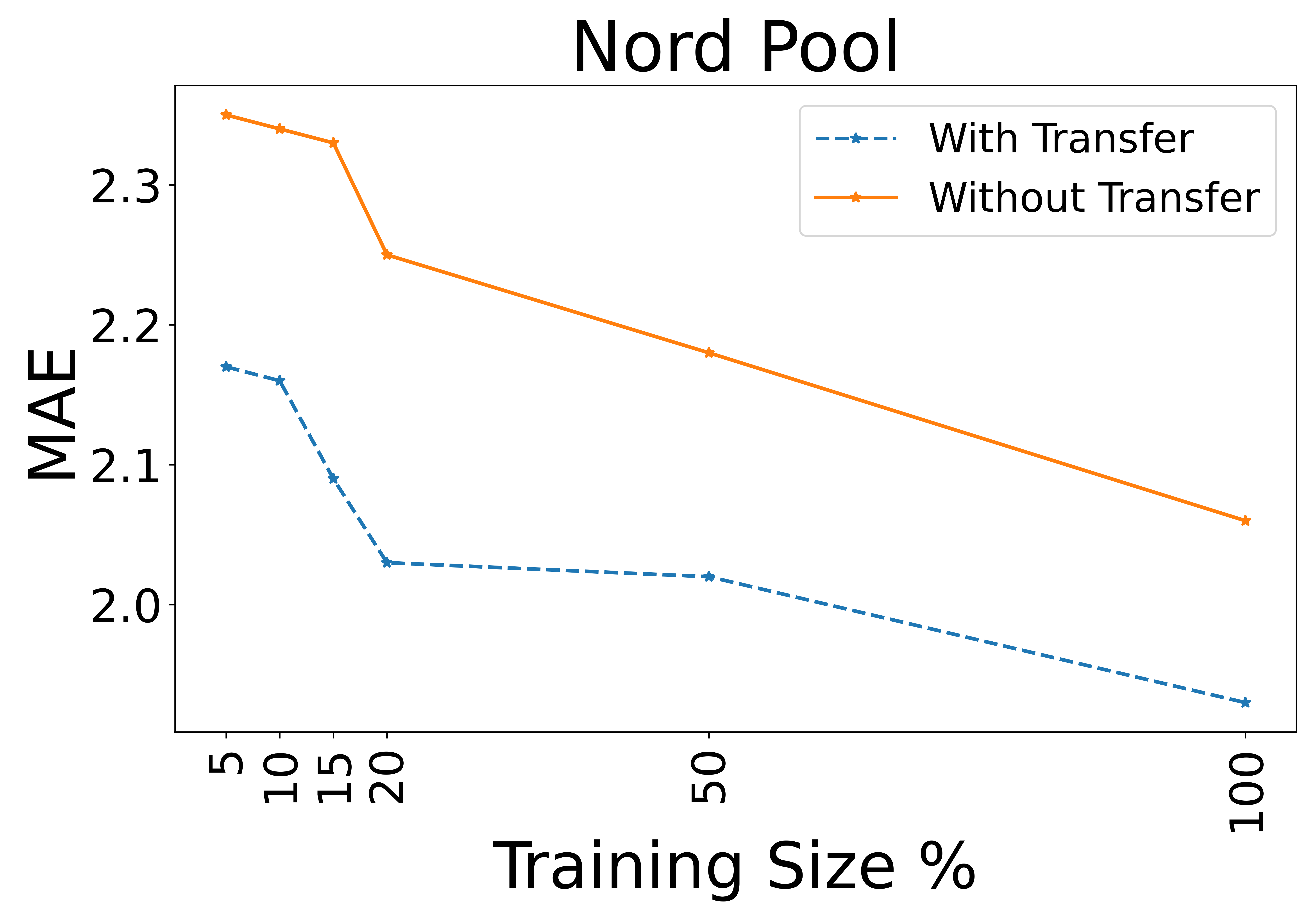}  
\end{tabular}
\caption{Transfer learning results for four markets using different amounts of training data. Transfer learning improves the performance according to various amounts of training data for all four markets compared to without transfer learning approach. The performance difference is more evident, when less training data is available.}
\label{fig:Transfer_result}
  \end{centering}
\end{figure}

We compare the performance of transfer learning with the single market training and report the results for each of the four markets in Figure \ref{fig:Transfer_result}. The training for transfer learning starts with the pre-trained network on other markets as described in Section \ref{sec:methods}. We use various amounts of training data in order to highlight the effect of transfer learning according to the availability of different amounts of data. Our results demonstrate that in all four markets, transfer learning improves the performance in comparison to single market training. Additionally, the performance increase is more significant, when less data is available for fine-tuning (e.g. re-training on target market).

\subsection{Diebold-Mariano Tests}
\label{sec:Dm_test}

We provide experimental results for the superiority of transfer learning in Table \ref{table:Quan} and the advantage of using less training data in transfer learning in Figure \ref{fig:Transfer_result}. Both of these evaluations illustrate a ranking between different cases, but statistical significance can not be assessed. In order to showcase the statistical significance, we use an adaptation of a multivariate variant of Diebold-Mariano (DM) test \cite{ziel2018day}.

\begin{figure}[ht]
\begin{tabular}{cc}
\includegraphics[width=0.47\linewidth]{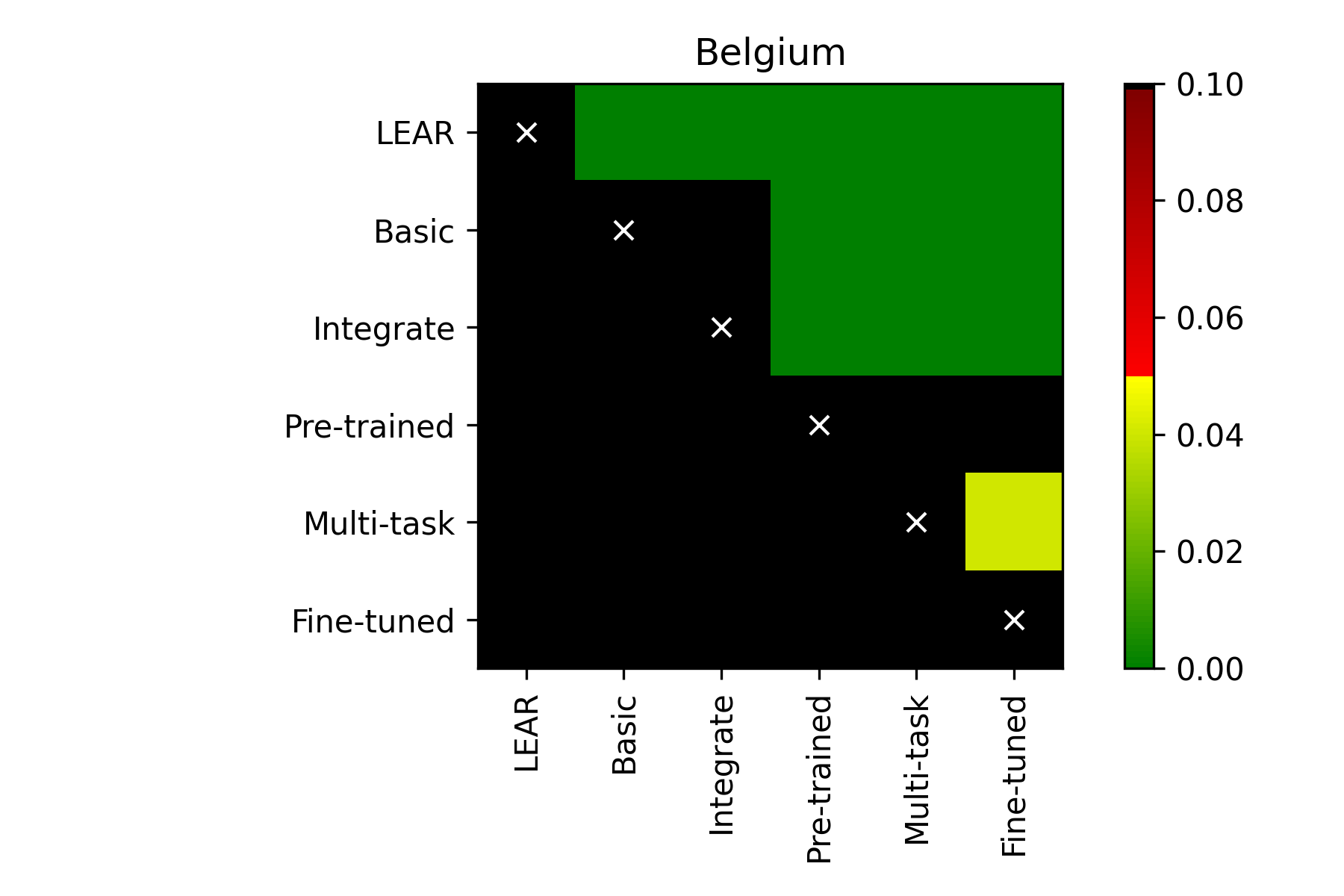} &  
\includegraphics[width=0.47\linewidth]{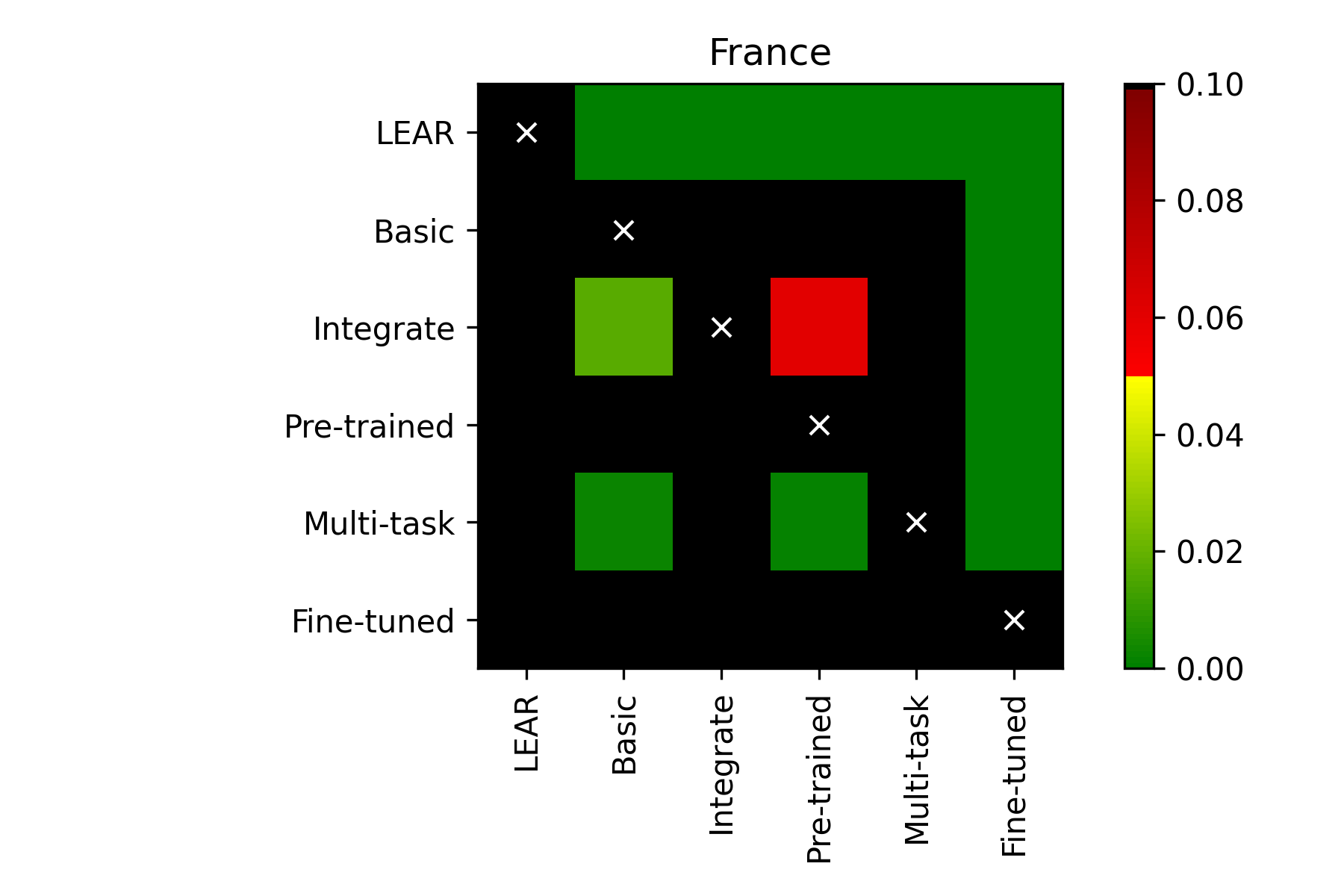} 
\end{tabular}

\begin{tabular}{cc}
 \includegraphics[width=0.47\linewidth]{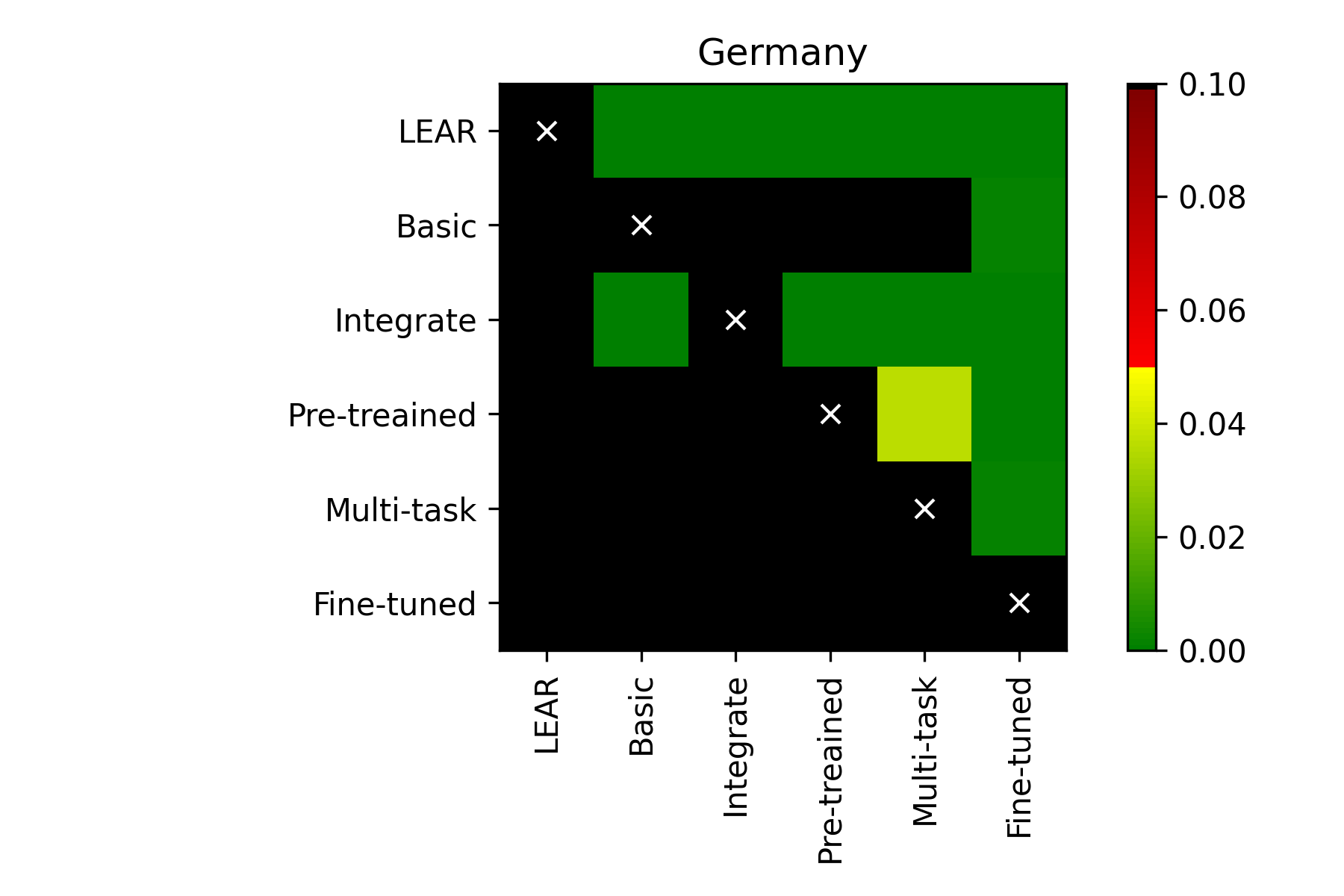}   & 
 \includegraphics[width=0.47\linewidth]{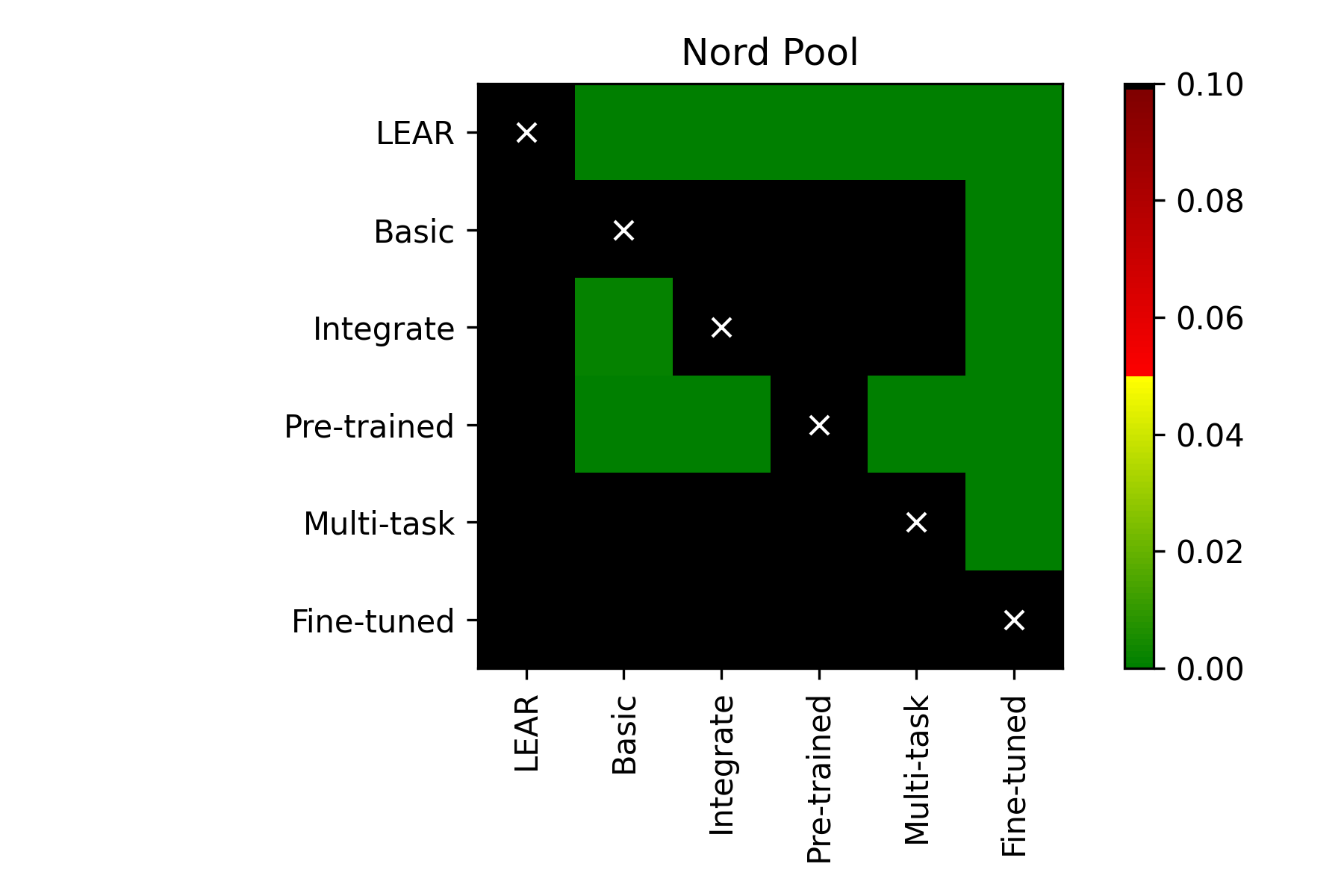}  
\end{tabular}
\caption{Diebold-Mariano tests for the results in Table \ref{table:Quan}. Green square demonstrates significantly better forecast of the model on the x-axis compared to model on the y-axis.}
\label{fig:DM_test1}
\end{figure}

\begin{figure}[!hb]
\begin{tabular}{cccc}
\includegraphics[width=0.21\linewidth]  {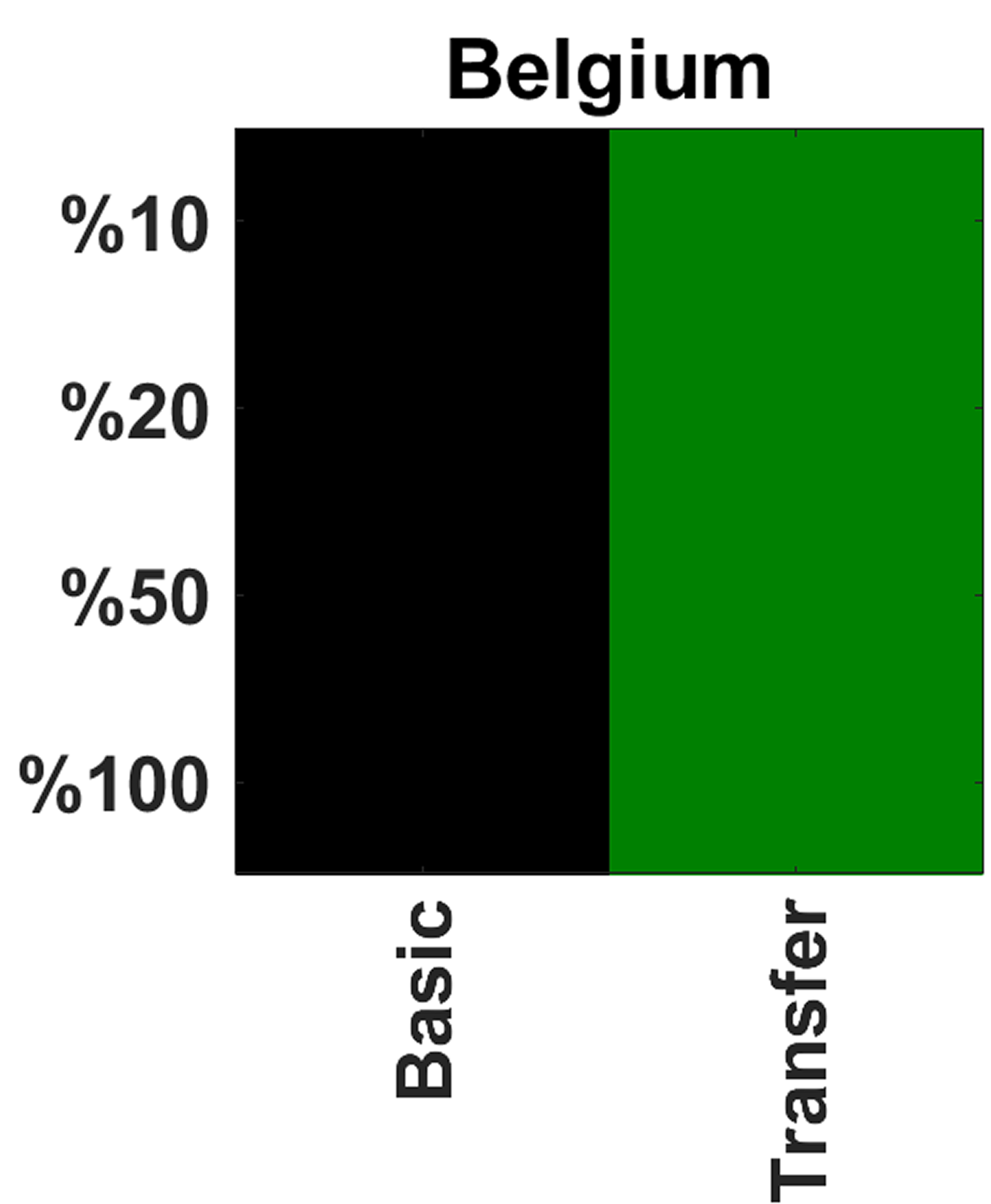} & 
\includegraphics[width=0.21\linewidth] {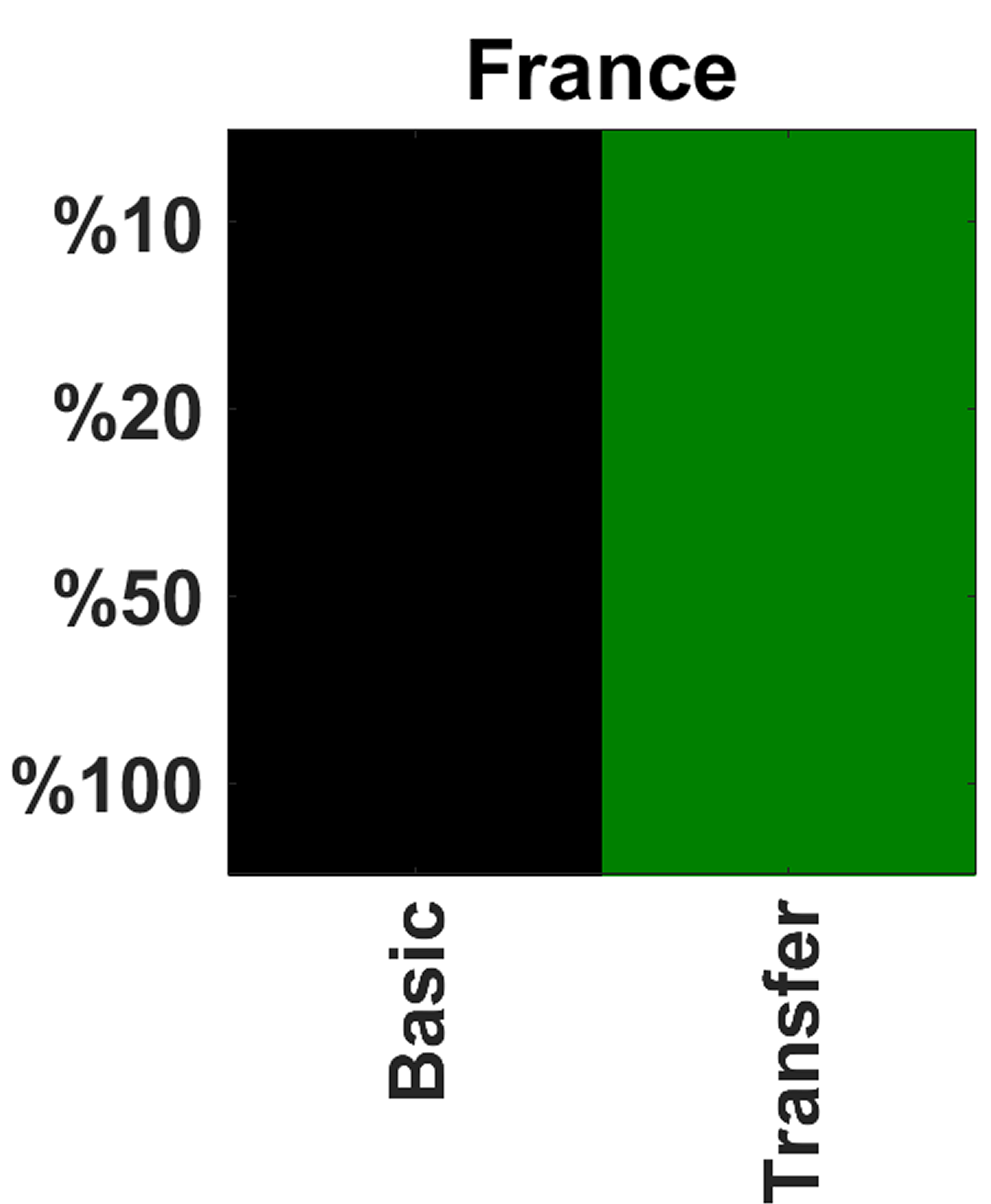} & 
 \includegraphics[width=0.21\linewidth] {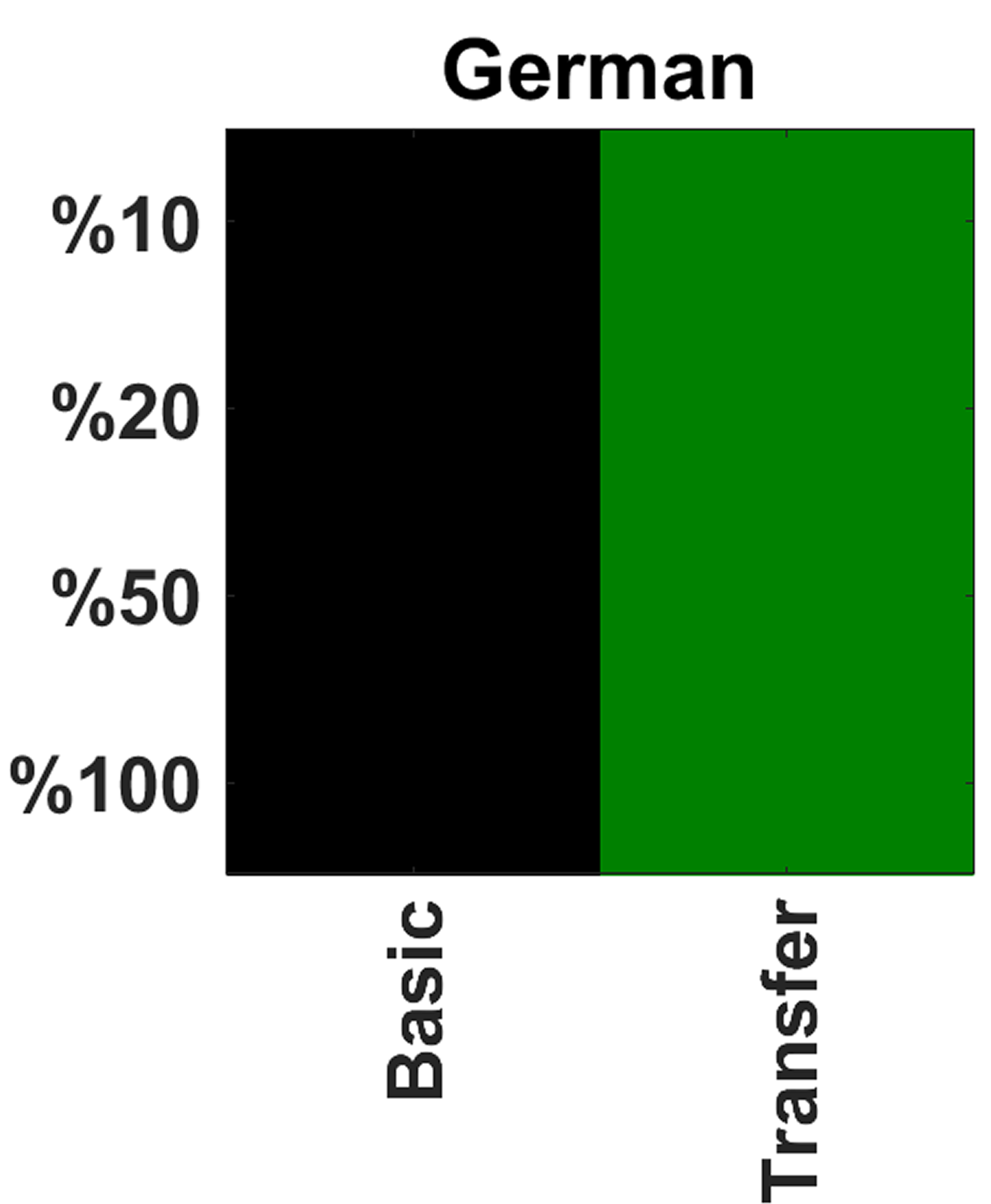} & 
 \includegraphics[width=0.26\linewidth] {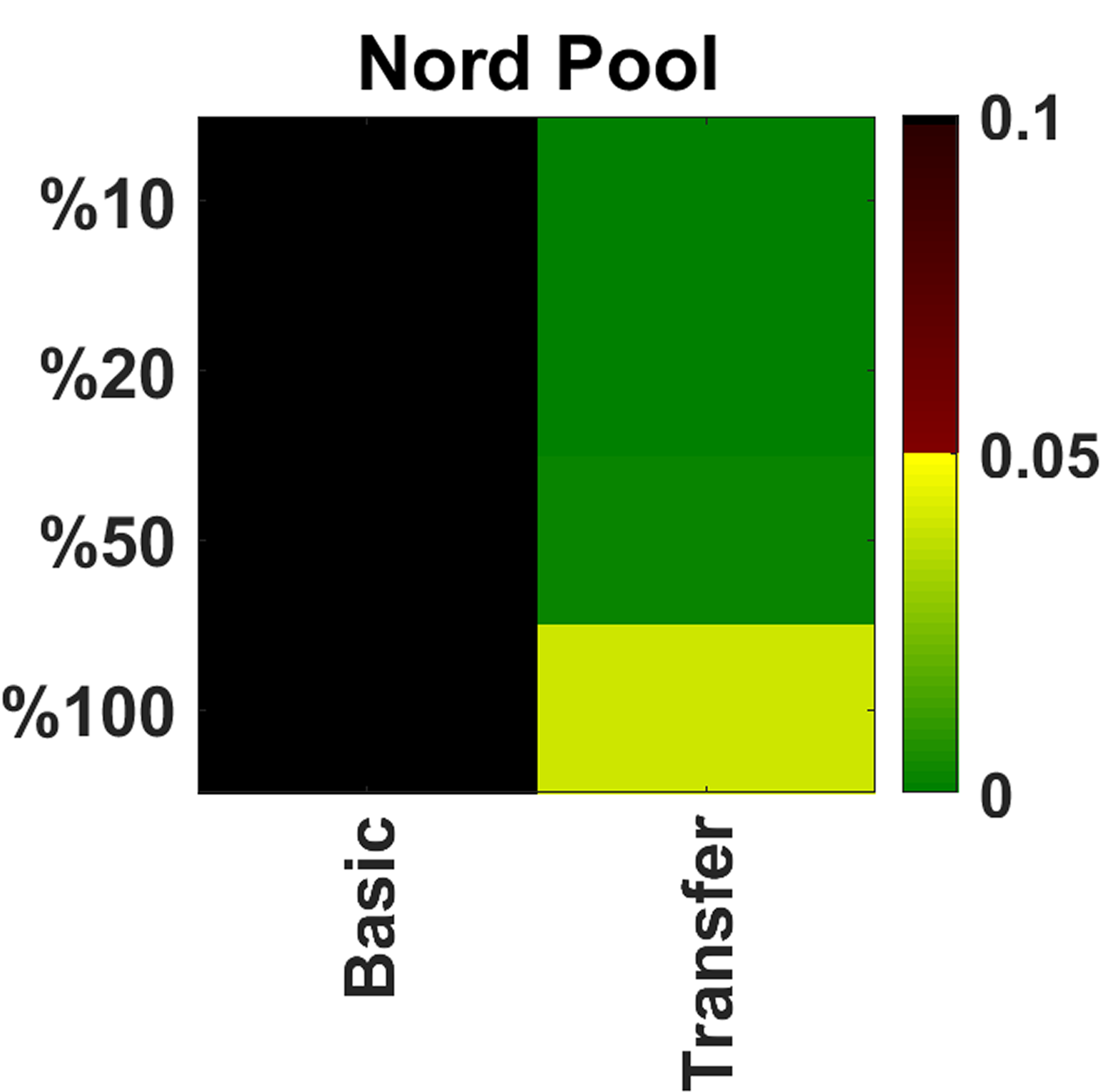}  
\end{tabular}

\caption{Diebold-Mariano tests for transfer learning with different sizes of fine-tuning data represented in Figure \ref{fig:Transfer_result}. Green square demonstrates significantly better forecast of the Transfer model to Basic model. Transfer learning showed statistically significant performance increase for all the cases.}
\label{fig:DM_test2}
\end{figure}

We use a multi-step ahead time series prediction approach \cite{cheng2006multistep} for forecasting the next day's 24 hours' prices. We develop a model to forecast the entire 24 prices of each day. Therefore, we have forecast values for the 2016 test data for each method (e.g. basic, integrate, pre-trained, multi-task, and fine-tuned). We compare the forecasts with the actual values and calculate the absolute error values for all models. Then, we apply the DM test and check whether one method is superior to the other method in statistically significant terms according to the one-tailed test. In Figure \ref{fig:DM_test1}, we show the p-values of the Diebold-Mariano test, between the different usages of data, for training neural networks. The tests are performed for each pair of training schemes and uses a color map to indicate the p-values. Statistically significant performance is presented with low p-values for the methods in x-axis versus y-axis. It is clear that fine-tuned outperforms other methods in a statistically significant way with the exception of pre-trained network in Belgian market. In Figure \ref{fig:DM_test2}, we illustrate the statistical significance of transfer learning on four markets for increasing number of training samples. For all portions of fine-tuning data from the target market, transfer learning generates statistically significant better results.

\section{Comparison With State of the Art Methods}
%

\label{sec:Exogenous}
In this section, we aim to compare our transfer learning idea on a rolling window scheme with state of the art methods. Data, exogenous variables and experimental design is different from the previous sections. We benefit from better exogenous variables and optimal models from the open-access toolbox \cite{lago2021forecasting} and report experiments in an open source repository \footnote{\url{https://github.com/salihgunduz/epftoolbox_transfer_learning}}.

\subsection{Data and Methods}

\begin{figure}
\begin{center}
    \includegraphics[width=\linewidth]{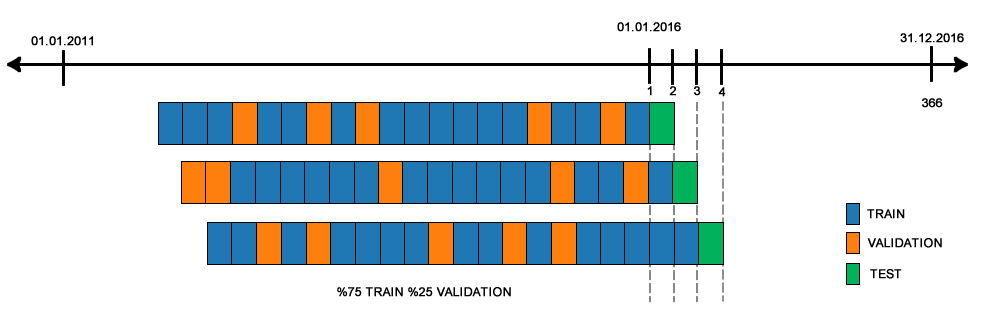}
    \caption{Training, validation and test parts for daily re-calibration.}
    \label{fig:germany2}
\end{center}
\end{figure}

We use data between 2011 and 2015 as the training and validation period(starting date of the training data varies for each market as reported in \cite{lago2021forecasting}) and 2016 as the test period (comply with our main research). We re-calibrate the models by using a rolling calibration window for each test day, which enables better forecasting performance with a new model for each day (Figure \ref{fig:germany2}). The experimental setup utilizes an open access-data set to use the best exogenous variables for each market \cite{lago2021forecasting}. All data before 2016 is divided into two-parts for training and validation. 75\% is used for training and 25\% is used for validation. In this experimental setup, training is re-done for every test day by moving the re-calibration window (rolling window scheme). 

The open-access data set contains different exogenous variables for each market except for BE and FR. BE and FR utilize “generation forecast” and “system load forecast” in FR. DE market has “wind power forecast” and “Ampirion zonal forecast” as exogenous inputs.  We perform transfer learning from Germany to France and Belgium to show contribution changes, when input features are diverse (Belgium and France have similar exogenous features, Germany has diverse features).

\subsection{Experimental Results}

In this section we perform two experiments to evaluate the influence of fine-tuning from different markets on the same model, and the influence of fine-tuning from different models on the same market. We use transfer to Belgium and French markets using France, Belgium and Germany as source markets. In the second experiment, we use different models for fine-tuning on single market (France).

\subsubsection{Fine-tuning on Different Markets}

\begin{figure}[ht]
\centering
\begin{tabular}{cccc}
\includegraphics[width=0.23\linewidth]{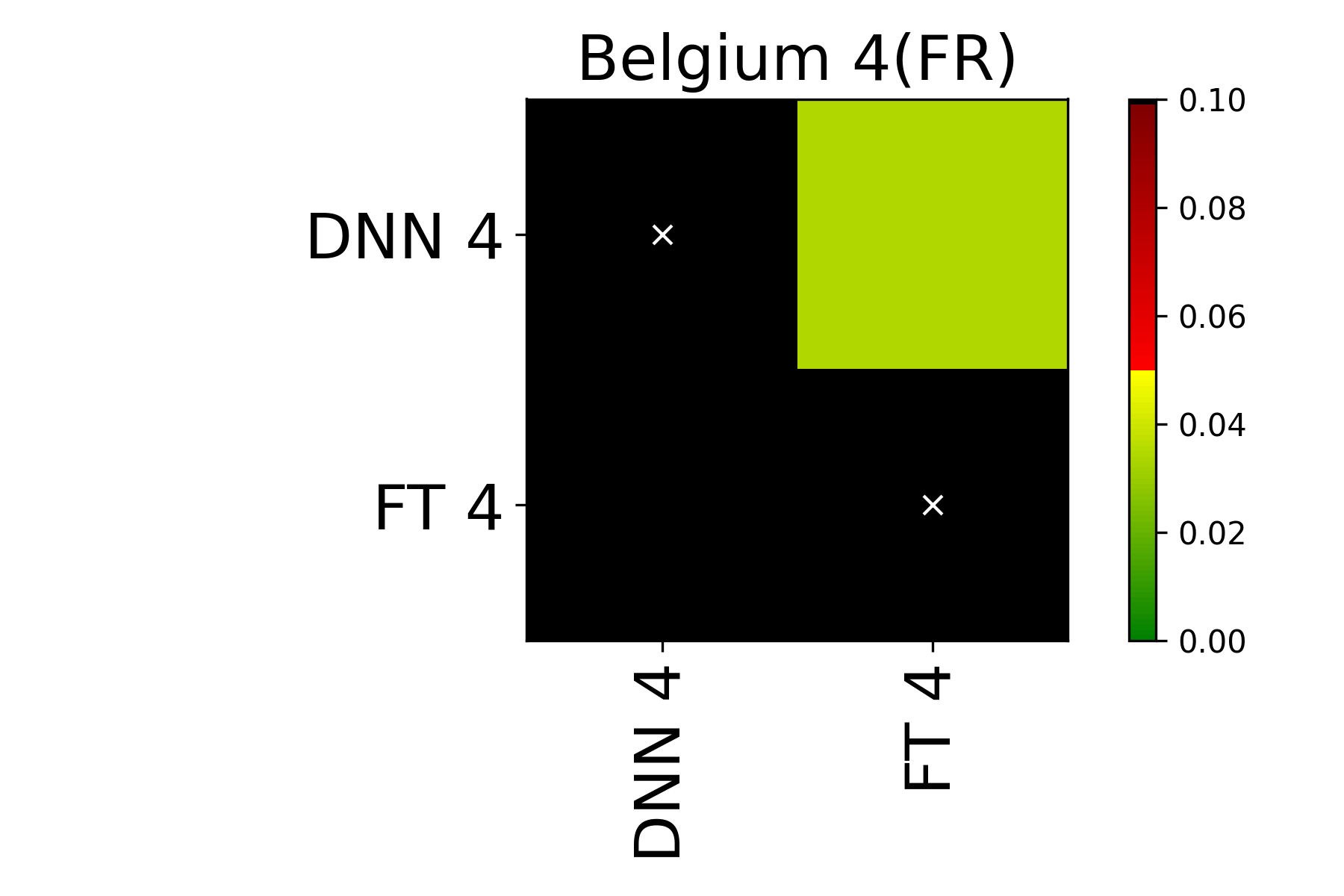} &  
\includegraphics[width=0.23\linewidth]{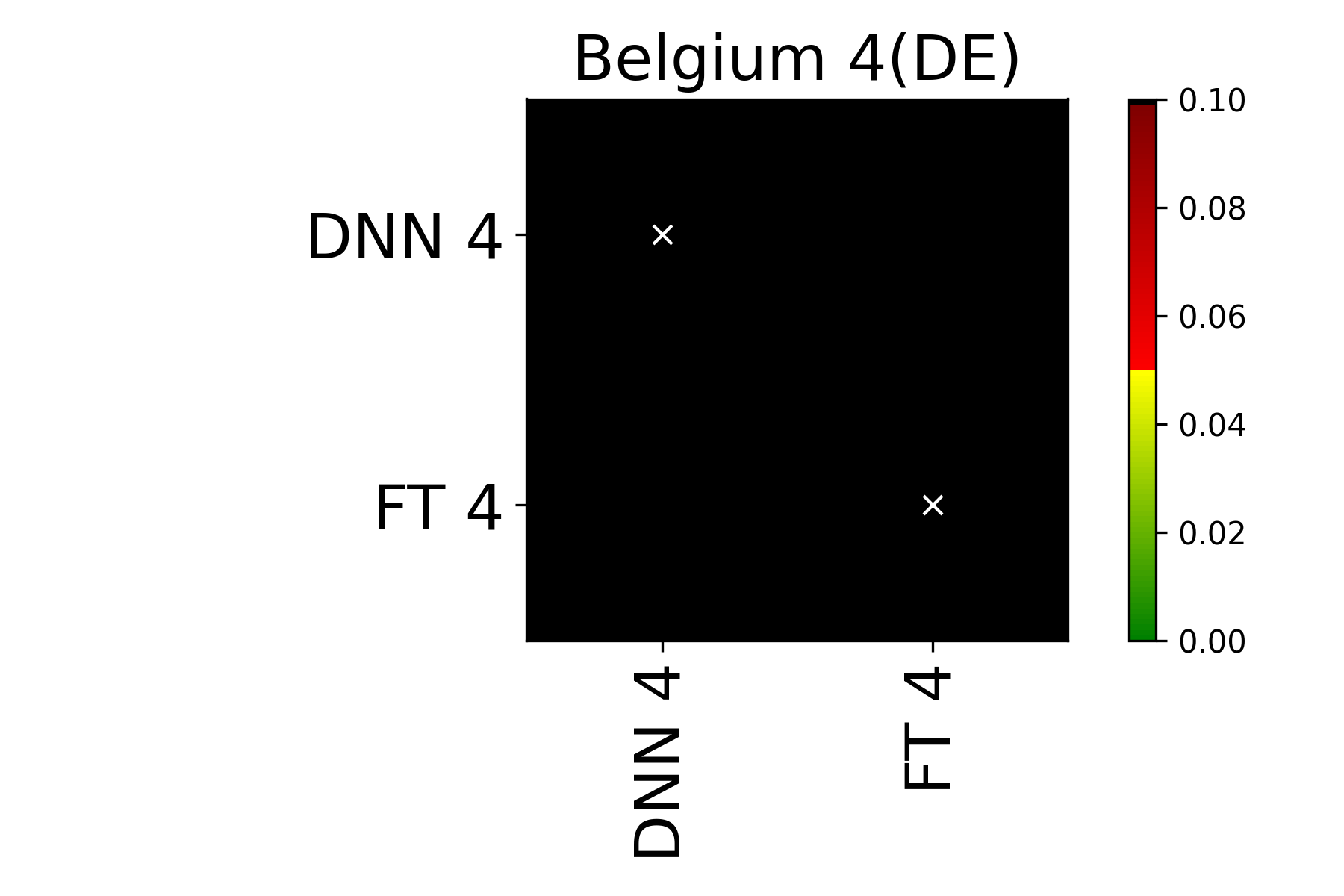} &
\includegraphics[width=0.23\linewidth]{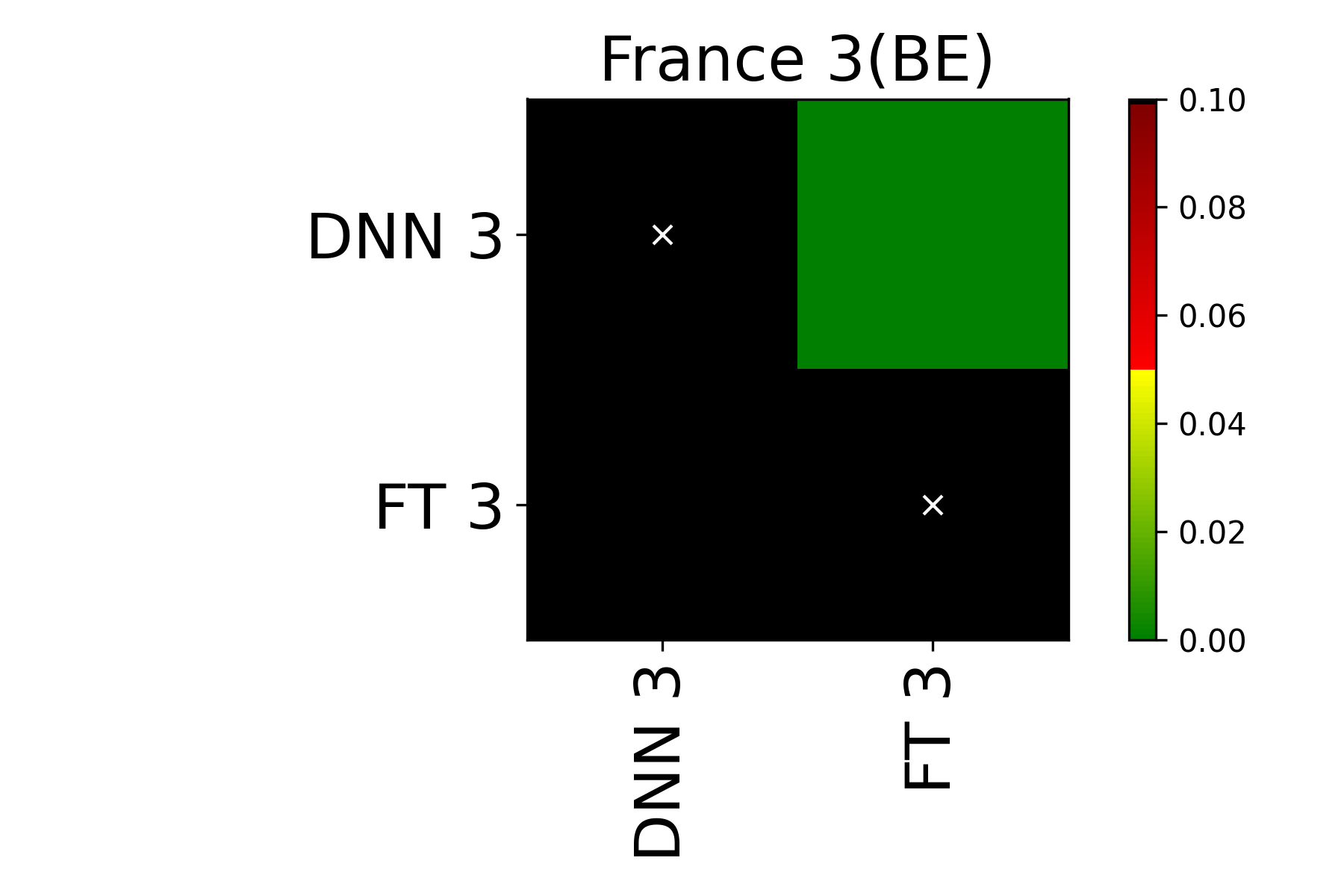}  & 
\includegraphics[width=0.23\linewidth]{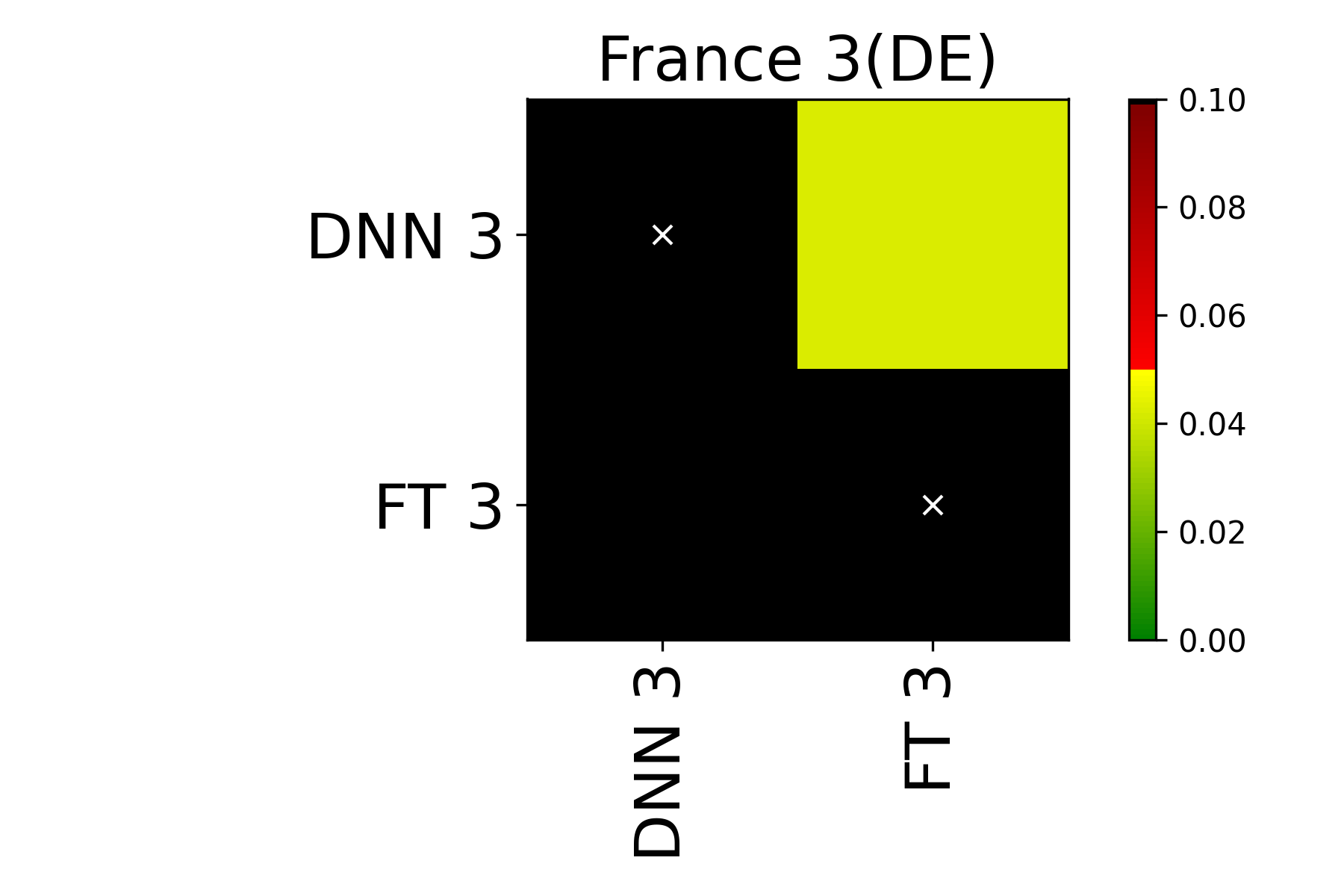}  
\end{tabular}
\caption{Diebold-Mariano tests for the results in Table \ref{table:FT_single}. Source markets are written in the parenthesis on the titles. For example, the first result is for the BE market. The model is pre-trained with FR and fine-tuned (FT) with Belgium data. Green square demonstrates significantly better forecast of the model on the x-axis compared to model on the y-axis.}
\label{fig:DM_single}
\end{figure}

First, we select the best DNN models for the Belgium and France target markets from \cite{lago2021forecasting} and report the results in Table \ref{table:FT_single} in comparison with fine-tuning results from different markets. We achieve improved results for fine-tuning in all four scenarios. Overall transfer between France and Belgium is generating better results when compared to using Germany as a source market. This conclusion is also reflected on DM tests in Figure \ref{fig:DM_single}.

\begin{table}
\centering
\caption{Open-access state-of-the-art models' transfer learning results for 2016. Experiments are performed with single-source market. Different source market's contributions are depicted.}
\begin{tabular}{ccccc}
    \hline
        Target & Source & Best Model                    & Basic (MAE) & Fine-tuned (MAE) \\ \hline
        FR     & BE     & \cellcolor[HTML]{FFFFFF}DNN 3 & 4.16        & \cellcolor[HTML]{C0C0C0}\textbf{3.87}    \\
        FR     & DE     & \cellcolor[HTML]{FFFFFF}DNN 3 & 4.16        & 4.02             \\\hline
        BE     & FR     & \cellcolor[HTML]{FFFFFF}DNN 4 & 4.97        & \cellcolor[HTML]{C0C0C0}\textbf{4.83}    \\
        BE     & DE     & \cellcolor[HTML]{FFFFFF}DNN 4 & 4.97        & 4.93    \\\hline
\end{tabular}
\label{table:FT_single}
\end{table}

\subsubsection{Fine-tuning on French Market with Different Models}

\begin{figure}[H]
\centering
\includegraphics[width=\linewidth*3/4]{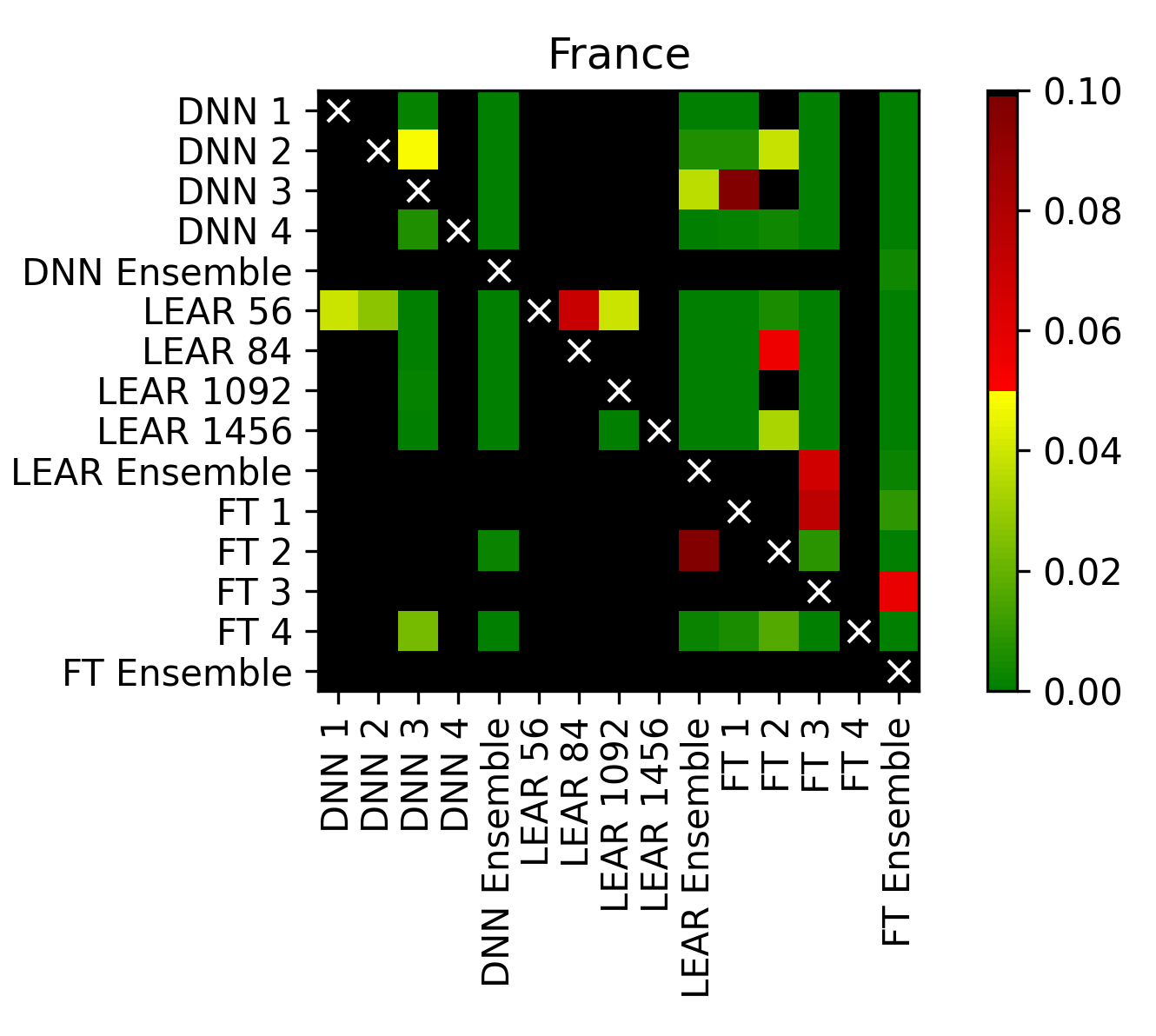}
\caption{Diebold-Mariano tests for the results in Table \ref{table:FT_France}. Data from source Belgium is used for pre-training the model. The pre-trained model parameters are utilized to initialize the fine-tuning step (re-training on the FR market). FT's are fine-tuned models.}
\label{fig:FR_all}
\end{figure}

We also selected the French market for detailed analysis on all DNN models (DNN 1 to 4) suggested in \cite{lago2021forecasting}. We applied transfer learning to all four models of the French market and results are reported in Table \ref{table:FT_France}.  Fine-tuned (FT) ensemble is the best model in terms of three metrics, where best performance in terms of RMSE is achieved by LEAR 84 method similar to \cite{lago2021forecasting}. The statistical significances of the model comparisons are demonstrated in Figure \ref{fig:FR_all}.  All fine-tuned methods (DNN 1 to 4) outperform their corresponding Basic DNN model in a statistically significant manner with the exception of Fine-tuned 4. These results indicate that fine-tuning is capable of improving the performance for various model designs.

\begin{table}
\centering
\caption{Comparison of the transfer learning approach with all models of the FR Market for 2016. Fine-tuning (FT) provides improved results for all DNN models (DNN 1 to 4) and FT ensemble (Ensembeling of 4 DNN models) generates best results in all metrics with the exception of RMSE.}
\begin{tabular}{lcccc}
    \hline
    \multicolumn{1}{c}{Model} & MAE           & rMAE          & sMAPE          & RMSE           \\\hline 
    DNN 1                     & 4.43          & 0.78          & 12.43          & 15.71          \\
    DNN 2                     & 4.38          & 0.77          & 11.59          & 16.56          \\
    DNN 3                     & 4.16          & 0.73          & 11.45          & 15.90          \\
    DNN 4                     & 4.48          & 0.79          & 12.15          & 16.52          \\
    DNN Ensemble              & 3.93          & 0.69          & 10.54          & 15.94          \\
    LEAR 56                   & 4.67          & 0.82          & 13.04          & 15.27          \\
    LEAR 84                   & 4.54          & 0.80          & 12.98          &\cellcolor[HTML]{C0C0C0} \textbf{13.83} \\
    LEAR 1092                 & 4.45          & 0.78          & 13.99          & 15.06          \\
    LEAR 1456                 & 4.57          & 0.81          & 14.59          & 15.14          \\
    LEAR Ensemble             & 4.01          & 0.71          & 11.40          & 14.08          \\
    FT 1                      & 4.01          & 0.71          & 11.27          & 14.31          \\
    FT 2                      & 4.24          & 0.75          & 11.28          & 16.73          \\
    FT 3                      & 3.87          & 0.68          & 10.55          & 15.47          \\
    FT 4                      & 4.47          & 0.79          & 12.04          & 15.95          \\
    FT Ensemble               & \cellcolor[HTML]{C0C0C0}\textbf{3.77} & \cellcolor[HTML]{C0C0C0}\textbf{0.67} & \cellcolor[HTML]{C0C0C0}\textbf{10.20} & 15.24    \\\hline     
\end{tabular}
\label{table:FT_France}
\end{table}

\section{Discussion and Conclusion}

\label{sec:discussion_conclusion}
We present a comprehensive study on effective ways of using data from various markets for electricity price forecasting. Moreover, we illustrate the superiority of transfer learning by using data from four different electricity markets. Our fundamental contribution in this paper is the use of transfer learning as a tool for accurate electricity price forecasting. Furthermore, we perform an investigation of different data usage within the DNN model for price forecasting and illustrate that utilizing data from multiple markets for pre-training is proven to be efficient. 

Pretrain-Finetune achieves significant performance improvements compared to its counterparts for different training sizes as illustrated in Section \ref{sec:Dm_test}. This finding is in line with the finding of Laptev et al. \cite{laptev2018reconstruction}, where they report similar performance improvement with transfer learning for a different time-series problem. Figure \ref{fig:DM_test2} highlights the statistical significance when different portions of data are used as training data for four markets. One additional finding, which is in line with Zhou et al. \cite{zhou2020transfer}, is that the performance gains with transfer learning is more evident, when less data is available as shown in Figure \ref{fig:Transfer_result}. Stacking markets multiplies data, so having more training samples makes the model avoid over-fitting as indicated in Perez et al. \cite{perez2017effectiveness}. Table \ref{table:Quan} indicates the superior performance on all four markets when data from different markets are used together in a transfer learning scheme compared to the basic model. Moreover, we investigate the importance of including exogenous variables on transfer learning (e.g. temperature, day of the week) and illustrate improved performance compared to using only lagged price values. 

One additional finding is the improved performance of the models with re-calibration for each day and selection of appropriate exogenous variables. The improvement of fine-tuning can be observed in various model and exogenous variable setups. With limited tests on French, Belgian and German markets, it is observed that using the similar exogenous variables can aid having statistically significant performance increase. However, further analysis on input-output relationships (mappings) of the markets is required to understand the true contribution of transfer learning under different source-target market scenarios. 

One avenue of improvement for this work is the addition of multiple features to the training scheme (e.g. reserve margin). Novel deep learning frameworks and loss functions can be investigated to further boost performance. Using hybrid methods in this framework can also improve forecast performance. In the future, we aim to use our model for continuous learning and prospective price prediction. Transfer learning enables real-time forecasting, where the model is not necessarily trained from scratch at each prediction and previous data is utilized in an efficient framework. We also believe the investigation of similar techniques can be instrumental in intraday markets, where more trading data is available.

In conclusion, we demonstrate that transfer learning can be used as an efficient tool for electricity price forecasting. This approach can be applied to suitable markets without requiring a large amount of training time \cite{tian2019similarity}. Evidence from four markets shows that neural network-based models' ability and generalization capability make them a suitable choice for being used in electricity price forecasting.

\section*{Data Access Statement}
All data utilized in this research are publicly available. Price and exogenous variable data are downloaded from the following websites.

\url{https://transparency.entsoe.eu/}
\\\url{https://data.open-power-system-data.org/time_series/}
\\\url{https://data.open-power-system-data.org/weather_data/}

\bibliography{mybibfile}
\

\end{document}